%% file: sigir2023.tex
\documentclass[sigconf]{acmart}

\input{commands}

\AtBeginDocument{%
  \providecommand\BibTeX{{%
    \normalfont B\kern-0.5em{\scshape i\kern-0.25em b}\kern-0.8em\TeX}}}

\copyrightyear{2023} 
\acmYear{2023} 
\setcopyright{rightsretained} 
\acmConference[SIGIR '23]{Proceedings of the 46th International ACM SIGIR Conference on Research and Development in Information Retrieval}{July 23--27, 2023}{Taipei, Taiwan}
\acmBooktitle{Proceedings of the 46th International ACM SIGIR Conference on Research and Development in Information Retrieval (SIGIR '23), July 23--27, 2023, Taipei, Taiwan}\acmDOI{10.1145/3539618.3591935}
\acmISBN{978-1-4503-9408-6/23/07}
\begin{document}

\title{Metric-agnostic Ranking Optimization}


\author{Qingyao Ai}
\affiliation{%
	\institution{Dept. CS\&T, Tsinghua University\\Zhongguancun Laboratory}
	\city{Beijing} 
	\country{China}
}
\email{aiqy@tsinghua.edu.cn}

\author{Xuanhui Wang}
\affiliation{%
	\institution{Google Research}
	\city{Mountain View} 
	\state{CA} 
	\country{USA}
}
\email{xuanhui@google.com}

\author{Michael	Bendersky}
\affiliation{%
	\institution{Google Research}
	\city{Mountain View} 
	\state{CA} 
	\country{USA}
}
\email{bemike@google.com}

\renewcommand{\shortauthors}{Ai et al.}

\begin{abstract}
  Ranking is at the core of Information Retrieval.
  Classic ranking optimization studies often treat ranking as a sorting problem with the assumption that the best performance of ranking would be achieved if we rank items according to their individual utility.
  Accordingly, considerable ranking metrics have been developed and learning-to-rank algorithms that have been designed to optimize these simple performance metrics have been widely used in modern IR systems.
  As applications evolve, however, people's need for information retrieval have shifted from simply retrieving relevant documents to more advanced information services that satisfy their complex working and entertainment needs.
  Thus, more complicated and user-centric objectives such as user satisfaction and engagement have been adopted to evaluate modern IR systems today.
  Those objectives, unfortunately, are difficult to be optimized under existing learning-to-rank frameworks as they are subject to great variance and complicated structures that cannot be explicitly explained or formulated with math equations like those simple performance metrics. 
  This leads to the following research question -- how to optimize result ranking for complex ranking metrics without knowing their internal structures?
  To address this question, we conduct formal analysis on the limitation of existing ranking optimization techniques and describe three research tasks in \textit{Metric-agnostic Ranking Optimization}: 
  (1) develop surrogate metric models to simulate complex online ranking metrics on offline data;
  (2) develop differentiable ranking optimization frameworks for list or session level performance metrics without fine-grained supervision signals;
  and (3) develop efficient parameter exploration and exploitation techniques for ranking optimization in metric-agnostic scenarios. 
  Through the discussion of potential solutions to these tasks, we hope to encourage more people to look into the problem of ranking optimization in complex search and recommendation scenarios.
\end{abstract}

\keywords{Information Retrieval, Learning to Rank, Ranking Optimization}


\maketitle


\input{sec-intro}
\input{subsec-T1}

\input{subsec-T2}


\input{subsec-T3}

\input{sec-related}



\input{sec-broader}



\balance
\bibliographystyle{ACM-Reference-Format}
\bibliography{sigproc}

\end{document}

%% file: commands.tex
\usepackage{fancyhdr}
\usepackage{lscape}
\usepackage{latexsym}
\usepackage{amsfonts}
\usepackage{amsmath}
\usepackage{color}
\usepackage{epsfig}
\usepackage{xspace}
\usepackage{graphicx}
\usepackage{subfigure}
\usepackage{epsfig}
\usepackage{multirow}
\usepackage{url}
\usepackage{wrapfig}
\usepackage{enumitem}
\usepackage{caption}
\usepackage{epsf}
\usepackage{graphicx}
\usepackage{subfigure}
\usepackage{rotating}
\usepackage{mathtools}
\usepackage{mathrsfs}

\DeclareMathOperator*{\argmax}{arg\,max}
\DeclareMathOperator*{\argmin}{arg\,min}

%% file: sec-intro.tex

\section{Introduction}
\label{sec-intro}

Ranking is at the core of many Information Retrieval (IR) applications. 
Examples include but not limited to web page ranking on search engine result pages (SERPs), item ranking for recommendation, question/answer ranking for conversational search, etc.
Ranking performance affects not only the quality and efficiency of information access, but also high-level user experience such as engagement and satisfaction. 
Therefore, ranking optimization has long been considered as a key component of modern IR systems and a core problem in IR research community.

Classic ranking optimization studies often treat ranking as a sorting problem. 
Based on the Probability Ranking Principle (PRP)~\cite{robertson1976relevance}, they assume that the utility of a ranking candidate can be quantified as a scalar, and the quality of ranking can be maximized by showing results following the decreasing order of their utility scores.
This assumption essentially indicates that ranking performance can be measured by the individual utility of items on each position (with or without weighting schemes), and ranking tasks can be formulated as a set of partial order prediction problems. 
Therefore, learning-to-rank algorithms that build machine learning (ML) models to directly optimize training loss constructed from the weighting schemes of such ranking metrics can achieve the best performance in theory and have been shown to be effective in many ranking applications.
For simplicity, we refer to the ranking metrics built from PRP as \textit{simple ranking metrics} and learning-to-rank methods designed for them as \textit{simple ranking optimization} algorithms.

As applications evolve, however, recent studies have found a significant gap between the performance scores computed with simple ranking metrics and the actual ranking quality perceived by users~\cite{yilmaz2014relevance,10.1145/1277741.1277902,10.1145/2854946.2855005}.
In practice, users do not evaluate result utility independently in ranking following the PRP assumption. 
Their satisfaction on rankings is often affected by complicated factors such as the relations between results, the local context of the request, personal preferences of the users, etc.
Thus, research has been conducted in IR to provide better understanding on user behaviors in online systems, which results in numerous user-centric evaluation metrics such as satisfaction scores~\cite{chen2017user,liu2018satisfaction} and engagements~\cite{lagun2014towards,khandelwal2021jointly,zhao2020maximizing}.
We refer to these as \textit{complex ranking metrics}.

Despite the extensive studies on ranking evaluation in complex interaction scenarios, the techniques to support ranking optimization for complex ranking metrics are very much under development.
Unsurprisingly, it is non-trivial, if not impossible, to apply existing simple ranking optimization techniques to optimize complex ranking metrics. 
First, complex ranking metrics, especially those computed with user logs sampled from online systems, are subject to extensive noise produced by systematic bias or unpredictable user behaviors.
Because simple ranking optimization algorithms assume that each item have unique and fixed utility following the PRP assumption, the uncertainty in complex ranking metrics could easily confuse the algorithms and produce unreliable models.
Second, most existing complex ranking metrics are constructed on list levels instead of item levels. 
It is often difficult to decompose them into an aggregation of partial orders or pointwise losses on items, which are necessary for existing simple ranking optimization techniques.
Third, and most importantly, the modeling of complex ranking metrics itself is an unsolved research problem in IR and it is widely believed that it is impossible to derive a single mathematical framework to explicitly describe or interpret user-related ranking metrics~\cite{mao2016does}. 
In other words, the structures of complex ranking metrics could be \textit{agnostic}.
This is unacceptable in existing ranking optimization frameworks as they need explicit loss functions constructed based on target ranking metrics and the predicted utility (i.e., the ranking score) of each item in order to apply well-established algorithms (e.g., gradient descent) to optimize model parameters.
This leads to the following question:
\textit{How to optimize ranking without knowing the internal structures of ranking metrics?}

We believe that the answer to this question, which we refer to as \textit{Metric-agnostic Ranking Optimization}, would be to develop new sets of ranking theory and techniques on metric modeling, parameter optimization, and ranking construction. 
Those techniques would not only provide new ideas for ranking formulation, but also serve as the foundation of ranking optimization for complicated IR problems beyond the PRP assumptions and frameworks.
More specifically, to address metric-agnostic ranking optimization, we propose and discuss three key tasks in this paper, which are 
(1) \textit{structure-agnostic metric modeling and simulation}: how to build offline surrogate models for complex ranking metrics without extensive online experiments; 
(2) \textit{differentiable ranking process with coarse-grained reward}: how to make ranking functions or processes differentiable to coarse-grained reward on list or session level. 
and (3) \textit{efficient ranking exploration in parameter space}: how to conduct efficient and effective ranking exploration in parameter space without the PRP assumption.
Through the formulation and discussion of potential solutions to these tasks, we hope to inspire novel ideas and encourage more people to look into the problem of ranking optimization with complex ranking metrics. 

%% file: subsec-T1.tex
\vspace{-10pt}
\section{Structure-agnostic metric modeling and simulation}
\label{subsec-T1}

While there is no universal definition on what complex ranking metrics should look like, three characteristics are widely acknowledged to be key factors that make a ranking metric complex for ranking optimization: 
(1) complex ranking metrics are usually noisy and subject to large variance;
(2) complex ranking metrics are usually computed on list or session level, and 
(3) complex ranking metrics are often collected online with significant cost.
Take user engagement as an example.
The time people spend on a search/recommendation result pages varies greatly depending on their personalities.
This means that no universal threshold can be defined to distinguish good rankings from bad ones according to engagement time. 
Also, user engagement time on each result is often difficult to collect. 
In many cases, we only know how much time the user has spent on the whole result page but not those on each individual result.
Last but not least, while engagement time has been shown to be a valuable indicator of ranking performance, it's also expensive to get because it requires real users to interact with the online systems, which restricts both the size and the efficiency of data collection.


Due to these characteristics, direct optimization of complex ranking metrics is mostly impossible under existing learning-to-rank frameworks.
To the best of our knowledge, most existing studies on learning to rank formulate ranking as a sorting task where the best ranking is deterministic given the partial orders of item pairs.
The noisy nature of complex ranking metrics, however, indicates that the same ranking could receive different rewards in practice, which can easily confuse existing ranking optimization algorithms.
Also, existing learning-to-rank algorithms require explicit labels, either gathered from user behaviors or domain experts, on each individual result for partial order prediction. 
These labels are mostly unavailable for complex ranking metrics because the metric scores are computed on list or session level. 
Further, when ranking metrics are computed on list levels, the training data built from online or offline search logs would be extremely sparse as observed data are only a tiny portion of all possible rankings.


To address these problems, we need to bridge or alleviate the gap between the noisy and sparse nature of complex ranking metrics and the need of reliable and extensive training data for ranking optimization.
We believe that one promising direction is to construct surrogate models that can predict complex ranking metric scores on any rankings without online experiments or hypothesizing on metric structures.
Formally, let $\mathcal{I}$, $I$, and $i$ be the universal set of ranking sessions, the set of items in a session, and an item to be ranked, respectively. 
Let $\pi(I)$ be a specific ranking of items $i\in I$. 
For a complex ranking metric $m(\pi(I))$, the goal of metric modeling is to construct a surro\textit{g}ate model $g(\pi(I), \theta^*)$ that could serve as a reliable approximation of $m(\pi(I))$ with parameter $\theta^*$:
\begin{equation}
	\theta^* = \argmin_{\theta}\sum_{I\in \mathcal{I}}\sum_{\pi(I) \in \mathcal{R}(I)} l\big(g(\pi(I)|\theta), m(\pi(I))\big)
	\label{equ:surrogate_loss}
\end{equation}
where $\mathcal{R}(I)$ is the universal set of all possible rankings for $I$, and $l$ can be defined as any types of errors function such as pairwise loss (i.e., for two rankings $\pi(I)$ and $\pi'(I)$, the loss is 0 if $m$ and $g$ have same preferences on them, and is positive  otherwise).
In the rest of this section, we describe how to solve this metric modeling problem offline.
Specifically, we start from proposing new data sampling/weighting scheme for noise and variance quantification, and then move to the development of context-aware surrogate models for offline metric prediction.
After that, we describe an offline reinforcement learning framework to enhance the generalizability of metric surrogate models for training data generation.

\vspace{-8pt}
\subsection{Noise and Variance Quantification} \label{sec:noise}
Noise and variance in ranking evaluation is a common problem caused by multiple factors such as restrictive annotation process~\cite{arabzadeh2021shallow,buckley2007bias}, bias in user behaviors~\cite{yue2010beyond,joachims2017unbiased,agarwal2019addressing}, etc. 
Existing studies usually apply two types of methods to deal with noise in ranking metrics. 
The first one is to build user hypothesis through lab or field studies and use them to explicitly construct metric models to explain and eliminate noise in ranking evaluation~\cite{joachims2002optimizing,craswell2008experimental}.
These methods are particular useful for specific types of noise, but their generalizability is limited as user hypothesis could be inaccurate or out-of-date.
The other type of methods is to build ML models and use large-scale training data to learn and calibrate ranking metrics~\cite{chuklin2013click}.
While these methods are attractive due to their theoretical robustness, they are extremely data hungry.
Unfortunately, none of these existing methods is applicable to metric modeling in metric-agnostic ranking optimization because we neither have enough knowledge about the metric structure nor large-scale training data with annotations. 


In fact, the problem of noise quantification with limited data is not unique to metric modeling. 
For example, recent studies on unbiased learning and data sampling has received significant attention in the IR community.
Though designed for simple ranking optimization scenarios (i.e., optimize ranking metrics built on PRP), these studies show the possibility of combining the modeling of data noise with the learning of ranking models for effective noise quantification and reduction with limited data.
For instance, previous studies on unbiased learning to rank~\cite{ai2018unbiased,yang2020analysis} show that effective user bias and examination propensity models can be automatically built together with ranking models on search logs without online user experiments. 
More recently, studies on learning to rank with partial labeled data~\cite{cai2022negative} also show that by combining the learning of negative sampling strategies with ranking models, we can effectively discover noisy training samples in ranking optimization.

Based on these observations, we propose to explore the possibility of collaboratively learning noise quantification and surrogate metric models for complex ranking metrics.
Formally, given a complex ranking metric $m(\pi(I))$, we collect a set of training data $\mathcal{T}=\{\pi(I), m(\pi(I))|I \in \mathcal{T}_{\mathcal{I}}, \pi(I)\in T_I\}$ with ranking sessions $\mathcal{T}_{\mathcal{I}} \subset \mathcal{I}$ where each session contains multiple rankings $T_I$.
As discussed previously, $m(\pi(I))$ is not noise-free, and a direct quantification of noise in $m(\pi(I))$ is difficult due to the unknown structure of $m$ and limited data in $\mathcal{T}$.
Thus, instead of training a separate model to predict noise, we propose to construct a noise-\textit{o}bservation model $o(m,\pi(I))$ together with the surrogate model $g(\pi(I), \theta)$. 
The key observation is that a good noise-observation model can help us identify noisy data in training and improve the metric surrogate model, and a good metric surrogate model can help the noise-observation model better estimate noise in data.
Formally, let $(\pi^+\!(I), \!\pi^-\!(I))$ be a pair of rankings over $I$ where $m(\pi^+\!(I)) \!\!> \!\!m(\pi^-\!(I))$.
An example training loss of $g(\pi(I))$ (short for $g(\pi(I), \theta)$) can be defined as
\begin{equation}
\begin{split}
	L(g,\mathcal{T}) =  -\!\!\sum_{I \in \mathcal{T}_{\mathcal{I}}} ~~ \sum_{~~~~\pi^+(I),\pi^-(I) \in T_I}\!\!\!\!\!\!\!\!\!\!\log Pr(\pi^+,\pi^-|g),\\
	Pr(\pi^+,\pi^-|g) = \frac{e^{g(\pi^+(I))}}{e^{g(\pi^+(I))} + e^{g(\pi^-(I))}}
\end{split}
\label{equ:surrogate_g_loss}
\end{equation}
where we define $Pr(\pi^+,\pi^-|g)$ as the probability of $\pi^+(I)$ being better than $\pi^-(I)$ given $g(\pi(I))$.
Intuitively, when there is a higher probability of observing noise on $m(\pi^+(I))$ than $m(\pi^-(I))$, the reliability of $(\pi^+(I),\pi^-(I))$ should be discounted and reflected in the training of $g(\pi(I))$.
Inspired by counterfactual learning~\cite{joachims2017unbiased,ai2018unbiased}, we propose to revise Eq.~(\ref{equ:surrogate_g_loss}) by weighting data with 
\begin{equation}
	L(g,\mathcal{T}) \!\!=  \!\!-\!\!\!\!\sum_{I \in \mathcal{T}_{\mathcal{I}}} \!\!\big( \sum_{~~~~\pi^+(I),\pi^-(I) \in T_I}\!\!\!\!\!\!\!\!\!\!\! (1\!-\! Pr(\pi^+\!,\!\pi^-|o,m))\log Pr(\pi^+\!,\!\pi^-|g)~~~\big)
	\label{equ:surrogate_ipw_g_loss}
\end{equation}
where we define $Pr(\pi^+\!,\!\pi^-|o,m)\!\!=\!\! \frac{e^{o(m,\pi^+(I))}}{e^{o(m,\pi^+(I))}+e^{o(m, \pi^-(I))}}$ as the probability of observing inaccurate data on $(\pi^+(I), \pi^-(I))$, and use $1-Pr(\pi^+,\pi^-|o,m)$ to weight the data pair.
Similarly, we could also use $g(\pi(I))$ to revise our noise-observation model following the same weighting scheme and build training loss for $o(m,\pi(I))$ as
 \begin{equation}
 	L(o,\mathcal{T}) =  -\!\!\sum_{I \in \mathcal{T}_{\mathcal{I}}} \big(\!\! \sum_{~~~~\pi^+(I),\pi^-(I) \in T_I}\!\!\!\!\!\!\!\!\!\! (1-Pr(\pi^+,\pi^-|g))\log Pr(\pi^+\!,\!\pi^-|o,m) ~~~\big)
 	\label{equ:surrogate_ipw_o_loss}
 \end{equation}
where the probability of observing noise should be low when the surrogate model $g$ and the target metric $m$ agree with each other.
By iteratively optimize Eq.~(\ref{equ:surrogate_ipw_g_loss})\&(\ref{equ:surrogate_ipw_o_loss}), the two models help each other and efficiently extract noise and build surrogate model together. 

\vspace{-8pt}
\subsection{Metric Prediction with Local Context}
The nature of metric-agnostic ranking optimization creates significant difficulty in training data collection as we can only obtain metric scores on a whole ranked list that has been shown to users with no fine-grained feedback on each item.
The number of possible rankings are large in which only one or a tiny proportion of them has been shown to users. 
Since real-time ranking optimization with online user experiments are mostly prohibitive in practice, it's important to predict the metric scores, which we refer to as the \textit{ranking rewards}, for rankings without involving real users.
The most related studies on this direction are query performance prediction, which tries to predict the ranking performance of a retrieval system without actually showing the results to users~\cite{hauff2009combination,zhao2008effective}.
Methods proposed by these studies, however, are tailored for ad-hoc retrieval tasks~\cite{zhou2007query} and can only predict the performance of a query given a static ranking system.
This makes them inapplicable for our task as the rankings produced by ranking models are subject to significant variances during the training process.
In fact, how to predict the performance of a learning-to-rank system in general is an underexplored topic in the research community. 
 
In order to predict ranking rewards on unobserved rankings without online experiments, we need to construct metric surrogate models purely based on the data available in offline ranking optimization.
Among different learning-to-rank applications, one type of information is universally available in all ranking optimization scenarios, that is the \textit{local ranking context}.
Local ranking context refers to structure and feature information of a ranking and the items in it.
Given a reasonable ranking model, the local context of a ranking produced by this model often contains valuable information including session-specific feature distributions and inter-relations between items to rank. 
Previous studies on context-aware learning-to-rank~\cite{ai2018learning,ai2019groupwise,pang2020setrank} have shown that, by reading and utilizing the local context of an initial ranking, learning-to-rank models can better understand the meaning of relevance in current sessions and produce tailored rankings that better satisfies user's need.
We believe that such information could be a key for offline metric modeling in metric-agnostic ranking optimization.

In this task, we propose to construct offline surrogate models for complex ranking metrics with local ranking context.
Formally, given the training data $\mathcal{T}=\{\pi(I), m(\pi(I))|I \in \mathcal{T}_{\mathcal{I}}, \pi(I)\in T_I\}$, our goal is to construct function $g$ to predict $m(\pi(I))$ for any $\pi(I) \notin T_I$ with the local context of $\pi(I)$.
Let $\vec{x}_i$ be a feature vector (pre-computed with hand-crafted algorithms or learned together with ranking optimization) of item $i\in I$ in the current session, then the local context of $\pi(I)$ can be represented as $X_{\pi(I)}^g\!\!=\!\![\vec{x}_1, \vec{x}_2,...,\vec{x}_{|\pi(I)|}]$.
Based on Eq.~(\ref{equ:surrogate_loss}), the surrogate model $g$ can be learned by minimizing the empirical loss as:
\begin{equation}
	L(g,\mathcal{T}) = \sum_{I\in \mathcal{T}_{\mathcal{I}}}\sum_{\pi(I) \in T_I} l\big(g(X_{\pi(I)}^g|\theta), m(\pi(I))\big)
\label{equ:surrogate_context_training_loss}
\end{equation}
where $\theta$ is the model parameter of $g$.
The key idea is to use the local context $X_{\pi(I)}$ as inputs to $g$ so that, for any $\pi(I)\notin T_I$, we can predict its ranking reward without information from online experiments.
Note that the design of $g$ could be fairly flexible.
It could be as simple as a linear regression or as complicated as a self-attention network~\cite{vaswani2017attention}.
Also, we can easily adapt the noise reduction framework proposed in Section~\ref{sec:noise} by jointly optimizing $g(X_{\pi(I)}^g|\theta)$ with a noise-observation model $o(X_{\pi(I)}^o|\theta_o)$ where $X_{\pi(I)}^o$ is a set of factors that potentially create noise in $m(\pi(I))$.



\vspace{-8pt}
\subsection{Offline Reinforcement Learning for Surrogate Metric Modeling}
The most straightforward method to optimize the parameters of surrogate metric function $g$ is to directly apply supervised learning algorithms using the loss functions proposed above.
Yet, the generalizability of a supervised learning algorithm highly depends on the quality and quantity of training data.
In practice, numerous training data may not be available due to the difficulty of collecting complex ranking metrics online.
This jeopardizes the reliability and generaliability of metric modeling with supervised learning.

Fortunately, recent advances on reinforcement learning could provide some new insights on how to solve the data sparsity problem.
Reinforcement learning is originally designed for ML applications where an agent policy is built to interact with an environment in order to obtain certain rewards pre-defined for the task.  
Similar to the problem of online experiments in IR, learning such agents by trial-and-error with real users is often detrimental to the system as it could significant affect user experience.
Thus, an appealing solution is to construct agents purely based on logged transitions collected from the experiences of a previous policy that has already interacted with the environment for a while, which is referred to as \textit{offline reinforcement learning}.
Offline reinforcement learning has attracted considerable attention in the ML community~\cite{buckman2020importance,xiao2021general,wu2021uncertainty}.
Dadashi et al.~\cite{dadashi2021offline} have conducted a theoretical analysis of offline reinforcement learning for pseudometric learning and show that it can help the agent achieve significantly better performance than traditional supervised learning methods.
This gives us new inspirations on how to enhance model generaliability with limited data.


Therefore, we propose to adapt offline reinforcement learning algorithms to improve the generaliability of surrogate metric modeling.
In general, the key to offline reinforcement learning is a distance function for action states/trajectories and a pseudometric defined over the distance function.
The intuitive idea is that the reward of an unobserved state-action pair should be similar to an observed state-action pair in the logged transactions if their distance is small.
Formally, let $(s,a)$ be a state-action pair with observed reward $r(s,a)$.
Let $\pi_{\omega}(s)$ be a policy parameterized by $\omega$ that takes the current state as input to predict the next action, and $Q_\theta(s,a)$ be the action value model parameterized by $\theta$ that tries to predict the true reward of $(s,a)$.
Assuming that there is a distance function $d(s_1,a_1;s_2,a_2)$ that measures the distance between two pairs $(s_1,a_1)$ and $(s_2,a_2)$, we define $d_\mathcal{T}(s,a)=\min_{(s',a') \in \mathcal{T}}d(s,a,s',a')$ as the distance function between $(s,a)$ to observed data in $\mathcal{T}$.
Then, following a popular actor-critic framework~\cite{dadashi2021offline,wu2021uncertainty}, we can learn $\pi_{\omega}(s)$ and $Q_\theta(s,a)$ together by sampling transitions $(s,a,r,s')\in\mathcal{T}$ (where $s'$ is the state after action $a$ on $s$) and optimizing 
\begin{equation}
\begin{split}
\text{(critic)} \min_{\theta}||&Q_\theta(s,a)-r(s,a) - \alpha_rQ_\theta(s',\pi_{\omega}(s')) \\&~~~~~~~~~~~~~~~~~~~~~~~~~~~~~~~~~~~~~~~~~- \alpha_c\mathcal{F}(d_\mathcal{T}(s',\pi_{\omega}(s')))||,\\
\text{(actor)} \max_{\omega}~&Q_\theta(s,\pi_{\omega}(s)) + \alpha_a\mathcal{F}(d_\mathcal{T}(s,\pi_{\omega}(s)))
\end{split}
\label{equ:actor-critic}
\end{equation}
where $\alpha_r$, $\alpha_c$, and $\alpha_a$ are hyperparameters, and $\mathcal{F}(d_\mathcal{T}(s,a))$ is a pseudometric function defined over the distance between $(s,a)$ and observed data in $\mathcal{T}$.
Possible definition of $\mathcal{F}$ could be~\cite{dadashi2021offline}
\begin{equation}
	\mathcal{F}(d(s_1,a_1;s_2,a_2)) = |r(s_1,a_1)-r(s_2,a_2)| + \alpha_r\mathbb{E}_{a'\sim\mathcal{U}_\mathcal{A}}d(s_1',a',s_2',a')
\label{equ:offline_RL_f}
\end{equation}
where $s_1'$ is the next state from $s_1$ after action $a'$, and $\mathcal{U}_\mathcal{A}$ is a uniform distribution over action space $\mathcal{A}$.

Specifically for metric-agnostic ranking optimization, we can adapt the framework of offline reinforcement learning and define $s$ as a specific ranking of items (i.e., $\pi(I)$) and $a$ as an action of flipping the positions of two adjacent items in $s$.
Further, we can replace $r(s,a)$ with $m(\pi(I),a)=m(\pi(I)')$ (where $\pi(I)'$ is the ranking after flipping action $a$ on state $\pi(I)$) and define $d(\pi(I)_1,a_1;\pi(I)_2,a_2)$ as
\begin{equation}
	d(\pi(I)_1,a_1,\pi(I)_2,a_2) = \text{Jaccard}_{flip}(\pi(I)_1', \pi(I)_2')
\label{equ:flipping_d}
\end{equation}
where $\pi(I)_1'$ and $\pi(I)_2'$ is the rankings after action $a_1$ and $a_2$ on $\pi(I)_1$ and $\pi(I)_2$, respectively; and $\text{Jaccard}_{flip}(\pi(I)_1', \pi(I)_2')$ is the Jaccard distance between $(\pi(I)_1',\pi(I)_2')$ when the only actions are flipping adjacent items.
With different sample and approximation techniques on pseudometric learning, we can directly construct $\mathcal{F}$ from the distance function above and learn a surrogate metric model $Q_\theta(s,a)=g(X_{\pi(I)'}^g|\theta)$ with the offline actor-critic algorithms described in Eq.~(\ref{equ:actor-critic}).
Such learning algorithms have potential to produce better surrogate metric model $g$ as it can effectively regularize the prediction of $g$ on unobserved rankings based on their similarities to observed $m(\pi(I))$.
Also, depending on how we implement the action policy $\pi_{\omega}(s)$ (which is essentially a ranking model), we could improve the generaliability of $g$ by letting it interact with different types of ranking models or even jointly optimize both the ranking model and the surrogate metric model together.

%% file: subsec-T2.tex
\vspace{-10pt}
\section{Differentiable Ranking with Coarse-grained Reward}
\label{subsec-T2}

Among all existing ML algorithms, gradient descent is one of the most popular methods that have been applied to a variety of parameter optimization problems, especially for deep learning~\cite{goodfellow2016deep}.
Yet, unlike other tasks such as regression and classification, gradient descent is not directly applicable to ranking due to the discrete parameter space of ranking problems. 
For example, when we rank items by sorting them with ranking scores, a tiny perturbation on the score of an item wouldn't change the ranking and corresponding reward if it does not change the partial order between items.  
Thus, how to handle the discrete parameter space of ranking problems and make ranking models differentiable is a key research problem in learning-to-rank studies~\cite{liu2009learning}.


While there are many well-established optimization methods~\cite{joachims2006training,burges2005learning,burges2006learning,cao2007learning} proposed for non-differentiable ranking problems, most of them are not applicable to model-agnostic ranking optimization. 
In existing literature, there are two types of methods to compute gradient from discrete rankings. 
The first one, which is directly built on the PRP assumption, is to decompose a ranking as a set of partial order relations where we can treat each pair of items as a separate training example and train ranking models for partial order predictions instead~\cite{joachims2006training,burges2005learning}.
The second one is to interpret sorting with a probability framework so that we can model the likelihood of creating the best ranking given certain ranking metrics and simply optimize ranking models with maximum likelihood estimation~\cite{xia2008listwise,ma2021learning}.  
Despite their different motivations and formulations, all these methods assume the accessibility of fine-grained partial-order labels or information about the structure of ranking metrics so that they build training losses and apply gradient descent accordingly.
Unfortunately, in metric-agnostic ranking optimization, whether the best rankings can be achieved by correctly predicting partial item orders is questionable.
And, in many cases, we have neither fine-grained labels on item pairs nor knowledge on how the best rankings should look like in each ranking session.

In this section, we discuss the possibility of constructing differentiable ranking and parameter optimization algorithms for metric-agnostic ranking optimization.
Specifically, we propose to start from three directions:
(1) Differentialize discrete and deterministic ranking states with random item flips;
(2) Smooth the distribution of ranking rewards with stochastic ranking paradigms built on different sampling strategies;
(3) Separate item scoring from item ranking and propose ranking modules that adaptively decide how to present items based on score distribution and target metrics. 

\vspace{-8pt}
\subsection{Differentializing Ranking States through Random Flips}\label{sec:random_flips}
In general, ranking metrics or rewards are non-differentiable mainly because of the ``loose'' connection between ranking model outputs and actual item rankings.
For simplicity, assuming that rankings are created by sorting items based on their ranking scores (i.e., we only need a scoring function to rank items). 
Then, no matter how we perturb item's ranking scores, the ranking of items, so as the ranking reward, would remain unchanged as long as the partial orders of their ranking scores are unchanged.
This makes it impossible to compute parameter gradients directly from the reward we observed on each ranking.

\begin{figure}[t]
	\includegraphics[scale=0.36]{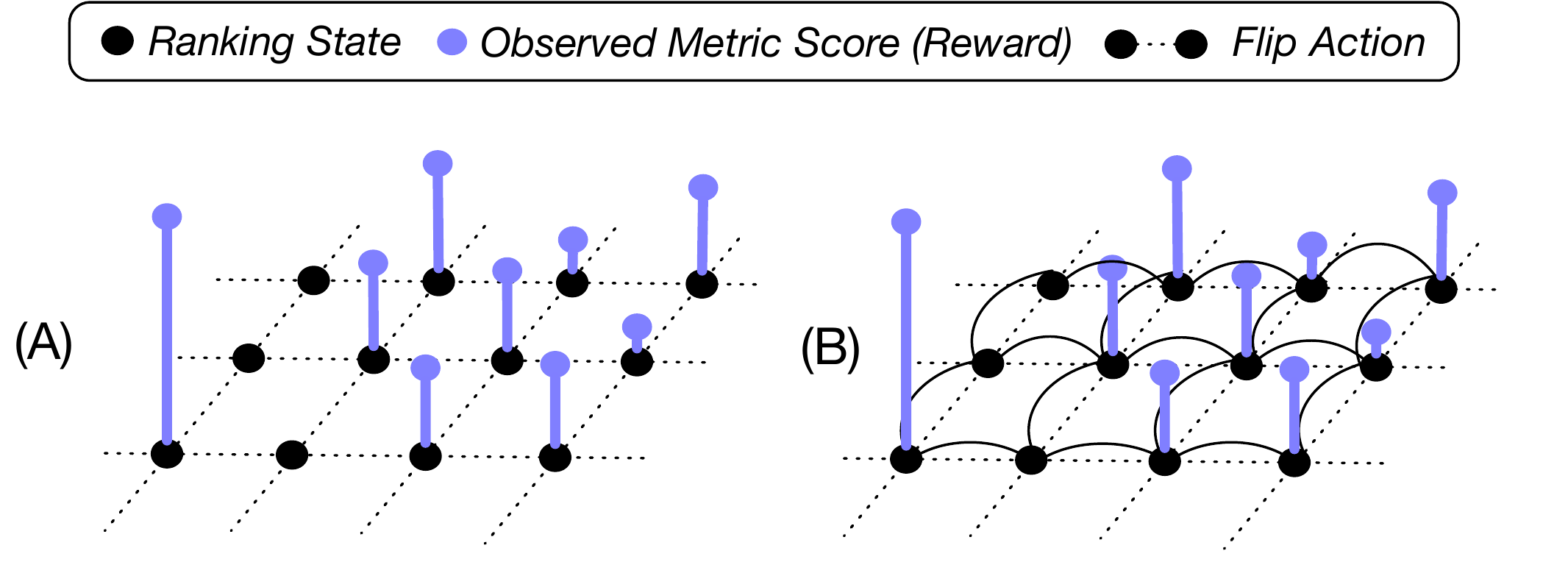}
	\caption{An illustration of latent ranking space.}
	\label{fig:ranking_state}
\end{figure}
More intuitively, let each possible item ranking be a state node in a latent space (Figure~\ref{fig:ranking_state}(A)). 
Let's define a \textit{flip} as swapping the positions of two items in a ranked list $\pi_a$, then the state $\pi_a$ is connected to a state $\pi_b$ if $\pi_a$ is same to $\pi_b$ after the flip.
Because the changes of ranking scores do not necessarily trigger a flip, the hyperplane of ranking metrics is non-differentiable with respect to the parameters of scoring functions.
Therefore, the key of differentiable ranking optimization is how to make flip actions sensitive to perturbations on ranking scores.
The most popular way to solve this problem in existing literature is to collect annotations on the partial orders of items so that we know how to flip items in advance and only need to train scoring functions to achieve it~\cite{burges2005learning}.
This method, however, is not applicable to metric-agnostic ranking optimization as the assumption that correct partial orders can guarantee best rankings (i.e., the PRP assumption) may not hold for complex ranking metrics, and we do not have access to partial order annotations on items in many cases.
Thus, new methodology and algorithms are needed for joint modeling flips and ranking scores.

We believe that the solutions of this problem lie in the modeling flip actions in ranking space and differentialization of ranking rewards on observed rankings.
Specifically, we propose to view flip actions as translation processes through noisy channels. 
Let $\pi_{ij}$ be the ranking of items when swapping the $i$th and $j$th items in ranking $\pi$, and let $m(\pi_{ij})$ be the observed metric score on $\pi_{ij}$.
We can define a noisy channel $\Gamma_{\pi_{ij}}\!\!=\!\!P_{\pi}(i|j)$ that translates the item on position $j$ to $i$ (i.e., swapping items on $i$ and $j$).
Then, for next translation on the $i$th item, the expected metric score $M(\pi_i)$ can be
\begin{equation}
	M(\pi_i) = \sum_{j\in \pi; \pi,\pi_{ij}\in \mathcal{T}}\frac{\Gamma_{\pi_{ij}}}{\sum_{k\in \pi, \pi_{ik}\in\mathcal{T}}\Gamma_{\pi_{ik}}}\cdot m(\pi_{ij})
\label{equ:noisy_reward}
\end{equation}
where $\mathcal{T}$ is the observed rankings in training data. 
Thus, given a set of observed rankings and corresponding metric scores, the goal of metric-agnostic ranking optimization is to construct a translation model that can translate any rankings into the ranking with highest reward.
There have already been mature methods to solve such problems in markov random fields~\cite{rue2005gaussian,metzler2005markov}, and we could easily adapt them to learn the noisy channel parameters. 
For instance, we can iteratively maximize the local expected value of $M(\pi_i)$ on all possible $\pi_i$ and stop when $M(\pi_i)$ converges. 
To connect the noisy channel $\Gamma_{\pi_{ij}}$ with ranking models, we could apply different types of noise models parameterized by the ranking scores of items on $i$ and $j$ (i.e., $s_i$ and $s_j$).
For example, we could assume that $\Gamma_{\pi_{ij}}$ follows a Gaussian Distribution $\mathcal{N}(\mu,\sigma)$ where $\sigma$ is a pre-defined variance and $\mu$ is computed as $(s_i-s_j)\cdot \text{sign}(i-j)$:
\begin{equation}
\Gamma_{\pi_{ij}} = P_{\pi}(i|j) = \int_{-\infty}^0 \mathcal{N}(x;(s_i-s_j)\text{sign}(i-j),\sigma) dx
\label{equ:noisy_channel}
\end{equation}
so that we can directly optimize $s_i$ by computing its gradient $\frac{\partial M(\pi_i)}{\partial s_i}$ through $\frac{\partial M(\pi_i)}{\partial \Gamma_{\pi_{ij}}}$ and  $\frac{\partial \Gamma_{\pi_{ij}}}{\partial s_i}$.
Note that the proposed noisy translation process is fundamentally different from existing studies that models ranking process as sequential item selection under probability frameworks~\cite{cao2007learning} as the former directly optimize ranking models based on observed data while the later requires the access of explicit metric scores on all possible rankings in order to compute the global  ranking losses.
In other words, the proposed method is particularly suitable for metric-agnostic ranking optimization in which we have limited knowledge on ranking metric and can only collect reward scores on limited number of rankings.

\vspace{-8pt}
\subsection{Stochastic Ranking Paradigms}
While ranking by sorting items with their ranking scores is a straightforward method, this paradigm oversimplifies ranking problems in practice and doesn't fit the needs of modern IR system today.
For example, item or user exposure fairness is considered to be a key problem for modern search and recommendation~\cite{singh2018fairness,biega2018equity}.
To satisfy users or balance content in different ethnic groups, retrieval results need to be ranked according to the distribution of user demographics and available information. 
Such demands cannot be satisfied with rankings created by sorting because a deterministic ranking will inevitably create uneven exposure to items in it.
Instead, recent studies on online learning to rank~\cite{oosterhuis2021computationally} and fairness~\cite{morik2020controlling} usually create multiple rankings for a set of items and sample  based on pre-computed data distributions to balance the exploration and exposure of items.
These methods provide a new perspective to view ranking problems and could potentially benefit our research on metric-agnostic ranking optimization. 

The main difference between rankings in fairness studies and existing learning-to-rank algorithms is a random process that could produce non-deterministic rankings over the same set of items, which we refer to as the \textit{stochastic ranking paradigm}.
Intuitively, stochastic ranking paradigms do not guarantee that the final rankings shown to users would strictly follows the partial orders of items determined by their ranking scores.
Different from the noisy channel modeling in Section~\ref{sec:random_flips}, stochastic ranking paradigms directly model the marginal probability of a ranking following a pre-defined probabilistic ranking generation model so that we can compute the prior distribution of rankings and sample them directly.
Thus, the key for ranking optimization with stochastic ranking paradigms is how to derive the prior distribution of ranking and how to compute or define loss functions on non-deterministic rankings.


Here, we propose to adapt stochastic ranking paradigms for metric-agnostic ranking optimization.
Formally, let $s=f(x|\theta)$ be the ranking score of an item $x\in I$ and the scoring function (i.e., the ranking model) $f$ is parameterized by $\theta$. 
Assuming that the metric scores of all possible rankings (i.e., $\{m(\pi)|\pi\in\mathcal{R}(I)\}$) are available beforehand, we can directly conduct metric-agnostic ranking optimization by maximizing the expected metric reward as
\begin{equation}
	\theta^*=\argmax_{\theta}\sum_{\pi\in\mathcal{R}(I)}m(\pi)\cdot Pr(\pi|f,\theta)
	\label{equ:true_stochastic_loss}
\end{equation}
where $Pr(\pi|f,\theta)$ is the prior probability of observing $\pi$ given the ranking generation model parameterized by $f(x|\theta)$.

There are two key advantages of stochastic ranking comparing to traditional rank-by-sort paradigms.
First, with the stochastic ranking process, we can compute the reward of a ranking policy without knowing the partial order labels on each item pairs.
Second, by parameterizing the ranking sampling distribution with the scoring function $f(x|\theta)$, we are able to compute the gradient of expected ranking reward with respect to $f$ and $\theta$ directly.
Specifically in practice, while the cost of collecting $m(\pi)$ for all possible $\pi$ is prohibitive, we can approximate $m(\pi)$ with the offline surrogate model $g(\pi)$ proposed in Section~\ref{subsec-T1} so that we can estimate rewards on unobserved rankings without online experiments.  
Further, for the modeling of prior ranking distribution $Pr(\pi|f,\theta)$, we could apply and test different probabilistic models.
For instance, we could adapt simple methods used in existing online learning-to-rank literature~\cite{tkachenko2016plackett,ma2021learning,oosterhuis2021computationally} (e.g., Plackett-luce distribution or multinomial logit), or try more advanced distributions such as CRS~\cite{seshadri2020learning}.
Also, instead of crafting $Pr(\pi|f,\theta)$ purely based on ranking scores $f(x|\theta)$, one can explore the possibility of incorporating rich model statistics such as scoring variance and output uncertainty in the construction of ranking models.
Previous studies have shown that uncertainty in ranking scores can be effectively estimated through Bayesian network~\cite{gal2016dropout} and is helpful for data exploration in online learning to rank~\cite{cohen2021not,yang2022can}.
We believe that such information could be useful for better modeling of prior ranking distribution and potentially benefit metric-agnostic ranking optimization. 



\vspace{-8pt}
\subsection{Detached Modeling for Scoring and Ranking}\label{sec:detach}
To the best of our knowledge, most existing studies on learning to rank~\cite{liu2009learning} does not explicitly differentiate the concept of \textit{scoring} and \textit{ranking}.
Existing literature usually refers to a ranking model as a model that generate ranking scores for each item to rank (i.e., $f(x|\theta)$ in T2.1), and, in the actual ranking process, they directly sort items with the scores or sample rankings based on pre-defined probabilistic models over scores.
For metric-agnostic ranking optimization, however, the integrated modeling of scoring and ranking has intrinsic disadvantages.
For instance, user's sensitivity to result relevance often changes depending on their information needs and application scenarios. 
It is possible that items must be strictly ranked by their relevance in some sessions (e.g., to guarantee user's efficiency in information seeking) or could be loosely ranked by their relative novelty in all candidates (e.g., to encourage user's exploration on results).
These diverse ranking needs cannot be satisfied with existing algorithms because they couple ranking with scoring and give almost no flexibility to the ranking process of items. 
When optimizing complex ranking metrics of which the internal structure could be unknown, such flexibility is important for building an effective ranking system.
 

To this end, we propose to detach the modeling of ranking with scoring in ranking problems.
Specifically, we can split the construction of a ranking model into two parts: a \textit{scoring module} that predicts item relevance based on their content and features; and a \textit{ranking module} that generate result rankings based on the scores produced by the scoring module. 
Let $s=f(x|\theta)$ be the ranking score of an item $x$, then we define a ranking function $\mathcal{H}(\{s\}|\rho)$ (parameterized by $\rho$) that generates a ranking $\pi(I)$ as
\begin{equation}
	\pi(I) \sim \mathcal{H}(\{f(x|\theta)|x\in I\}|\rho)
\label{equ:detached_ranking}
\end{equation} 
where $\mathcal{H}$ takes all ranking scores and context information as inputs to predict a ranking.
This separated formulation of scoring and ranking module creates a generic framework for ranking optimization as existing ranking paradigms can be treated as special cases in it, e.g., the sort-by-score paradigm~\cite{liu2009learning} can be viewed as a $\mathcal{H}$ implemented with a sorting algorithm; and the stochastic ranking paradigms with Plackett-luce~\cite{oosterhuis2018differentiable} can be seen as a $\mathcal{H}$ built with a Markov decision process and softmax functions.

The proposed ranking framework allows us to create separated models for the ranking module and thus introduce significant flexibility in terms of model design.
For instance, under a stochastic ranking framework, we can define $\mathcal{H}$ with a recurrent neural network and directly train the item sampling distribution based on the need of complex ranking metrics.
Specifically, we plan to explore collaborative learning algorithms for the training of $f(x|\theta)$ and $\mathcal{H}(\{s\}|\rho)$.
Given $\mathcal{H}(\{s\}|\rho)$, we can compute the prior probability of observing each $\pi(I)$ (i.e., $Pr(\pi|\mathcal{H}(\{f(x|\theta)|x\in I\}|\rho)$) and the expected metric reward with Eq.~(\ref{equ:true_stochastic_loss}).
Then, by iteratively fixing $\rho$ and $\theta$, we can jointly optimize $f(x|\theta)$ and $\mathcal{H}(\{s\}|\rho)$ as 
\begin{equation}
\begin{split}
\theta_t=&\argmax_{\theta}\sum_{\pi\in\mathcal{R}(I)}\!\!m(\pi)\!\!\cdot\!\! Pr(\pi|\mathcal{H}(\{f(x|\theta_{t-1})|x\in I\}|\rho_{t-1}) \\
\rho_t=&\argmax_{\rho}\sum_{\pi\in\mathcal{R}(I)}\!\!m(\pi)\!\!\cdot\!\! Pr(\pi|\mathcal{H}(\{f(x|\theta_{t-1})|x\in I\}|\rho_{t-1}) \\
\end{split}
\label{equ:scoring_ranking_optimization}
\end{equation}
where $\theta_t$ and $\rho_t$ are parameters for $f$ and $\mathcal{H}$ in time step $t$, respectively.
Further, we could consider factors that are important for complex ranking metrics but independent with item relevance (e.g., user personality) in the construction of $\mathcal{H}$ to further improve the performance of the generated rankings.
We believe that this detached modeling framework for scoring and ranking could have great potentials for metric-agnostic ranking optimization.


%% file: subsec-T3.tex
\vspace{-10pt}
\section{Efficient Ranking Exploration in Parameter Space}
\label{subsec-T3}

Efficiency and convergence rate are important factors to analyze the quality of an optimization algorithm.
It's particularly true for ranking optimization as the parameter space of ranking problems grows exponentially, if not faster, with respect to the number of items to rank.
To address this problem, previous studies on ranking optimization often decompose ranking problems into a set of partial-order classification tasks under the PRP assumption.
With fine-grained annotations on each partial-order pairs, this method can significantly reduce the parameter space of ranking problems from $O(n!)$ to $O(n^2)$ where $n$ is the number of items.
This makes it possible to optimize ranking models within polynomial time.
In metric-agnostic ranking optimization, however, there isn't any trivial solution to reduce the parameter space of ranking problems.
First, when  we possess limited knowledge on the structure of ranking metrics and metric scores can only be obtained on list or session level, the cost of collecting training data would be high and fine-grained annotations on each item or item pairs are difficult or impossible to get.
Second, due to their complicated nature, the reward hyper-surface of complex ranking metrics with respect to model parameters could have many local optima that hurt the efficiency and effectiveness of optimization algorithms. 


In this section, we discuss the development of efficient learning algorithms for metric-agnostic ranking optimization.
The goal is to reduce the parameter space or alleviate the need of data collection for complex ranking metric optimization and eventually make it possible to integrate the proposed metric-agnostic ranking optimization techniques into real IR systems.

\vspace{-8pt}
\subsection{Initialization with Multi-task Learning}
Building large-scale models from scratches is considered to difficult in many applications such as computer vision and natural language process. 
This is also true for metric-agnostic ranking optimization considering the numerous number of possible rankings and giant parameter space.
Previous studies~\cite{devlin2019bert} have found that pre-training models with large-scale unsupervised or weakly supervised data could significantly improve the efficiency and effectiveness of model optimization.
A key observation is that, through pre-training, we could better initialize model parameters and narrow down the parameter space to explore in further optimization.
Previous studies on learning to rank~\cite{ai2018unbiased,ai2018learning,yang2020analysis} also show that good parameter initialization or good initial ranker can guarantee the convergence and performance of ranking models both theoretically and empirically.

Inspired by these observations, one potentially promising direction to pursue is the exploration of different parameter initialization and regularization techniques for efficient metric-agnostic ranking optimization.
While the optimization of complex ranking metrics with unknown structures is still new to the IR community, there have already been many mature algorithms for how to construct training data and ranking models for simple metrics under PRP.
Therefore, we could construct a multi-task learning framework that combines the traditional ranking optimization algorithms with the proposed methods for complex ranking metrics.
Let $L_{C}(\mathcal{T}|\theta)$ be the loss function used by a metric-agnostic ranking optimization algorithm in Section~\ref{subsec-T2}
where $\mathcal{T}$ and $\theta$ are the training data and model parameters, respectively.
Let $L_{S}(\mathcal{T}'|\theta)$ be the loss function of an ranking algorithm for simple metric optimization (e.g., pairwise cross entropy over human-annotated labels).
Then, at time step $t$, we could update the model parameter $\theta$ as
\begin{equation}
	\theta_t \!=\! \theta_{t-1}\!+\!\alpha\big(\phi(t,\theta_{t-1})\frac{\partial L_{C}(\mathcal{T}|\theta)}{\partial \theta} + (1-\phi(t,\theta_{t-1}))\frac{\partial L_{S}(\mathcal{T}'|\theta)}{\partial \theta}\big)
\label{equ:multitask_optimization}
\end{equation}
where $\alpha$ is the learning rate and $\phi(t,\theta_{t-1})$ is a function that takes the historic model parameters or gradients to predict the current weights for different ranking losses.
The learning framework described in Eq.~(\ref{equ:multitask_optimization}) can be viewed as a multi-task learning algorithm that balance parameter optimization for two different targets.
The rational behind this is that, while the simple loss function $L_{S}(\mathcal{T}'|\theta)$ may not be perfect for complex ranking metrics, it should be able to produce a reasonably good starting point. 
By combining it with $L_{C}(\mathcal{T}|\theta)$, we allow the model to quick initialize parameters based on simple losses built on a separate or the same training data $\mathcal{T}'$ (with fine-grained labels) and thus increase the overall convergence rate of the optimization algorithms. 
Here, the weighting function $\phi(t,\theta_{t-1})$ is introduced to balance the parameter initialization process with the final objective in metric-agnostic ranking optimization.
It should give relatively high weights to $L_{S}(\mathcal{T}'|\theta)$ in the beginning and gradually move to $L_{C}(\mathcal{T}|\theta)$ when we have more confidence on the model parameters.
This could be achieved in multiple way such as implementing $\phi(t,\theta_{t-1})$ with a simulated annealing algorithm~\cite{van1987simulated} or a parameter-free convex learning algorithm (e.g. the coin betting algorithm using
Krichevsky-Trofimov Estimator~\cite{orabona2016parameter}). 
Also, $L_{S}(\mathcal{T}'|\theta)$ is not necessarily a single loss function and could be a combination of multiple loss functions built with different types of weak supervision signals~\cite{eiron2003analysis,dehghani2017neural}.

\vspace{-8pt}
\subsection{Exploration with Uncertainty Modeling}
Generally, large-scale training data are important, if not essential, for the generalizability and reliability of learning-to-rank models.
Unfortunately, in metric-agnostic ranking optimization, the scores of complex ranking metrics have to be collected online and it's often prohibitive to get data that are large enough to cover multiple rankings in each session.
Therefore, how to collect training data wisely is particularly important for the efficiency and convergence of ranking optimization algorithms.

\begin{figure}[t]
	\includegraphics[scale=0.1]{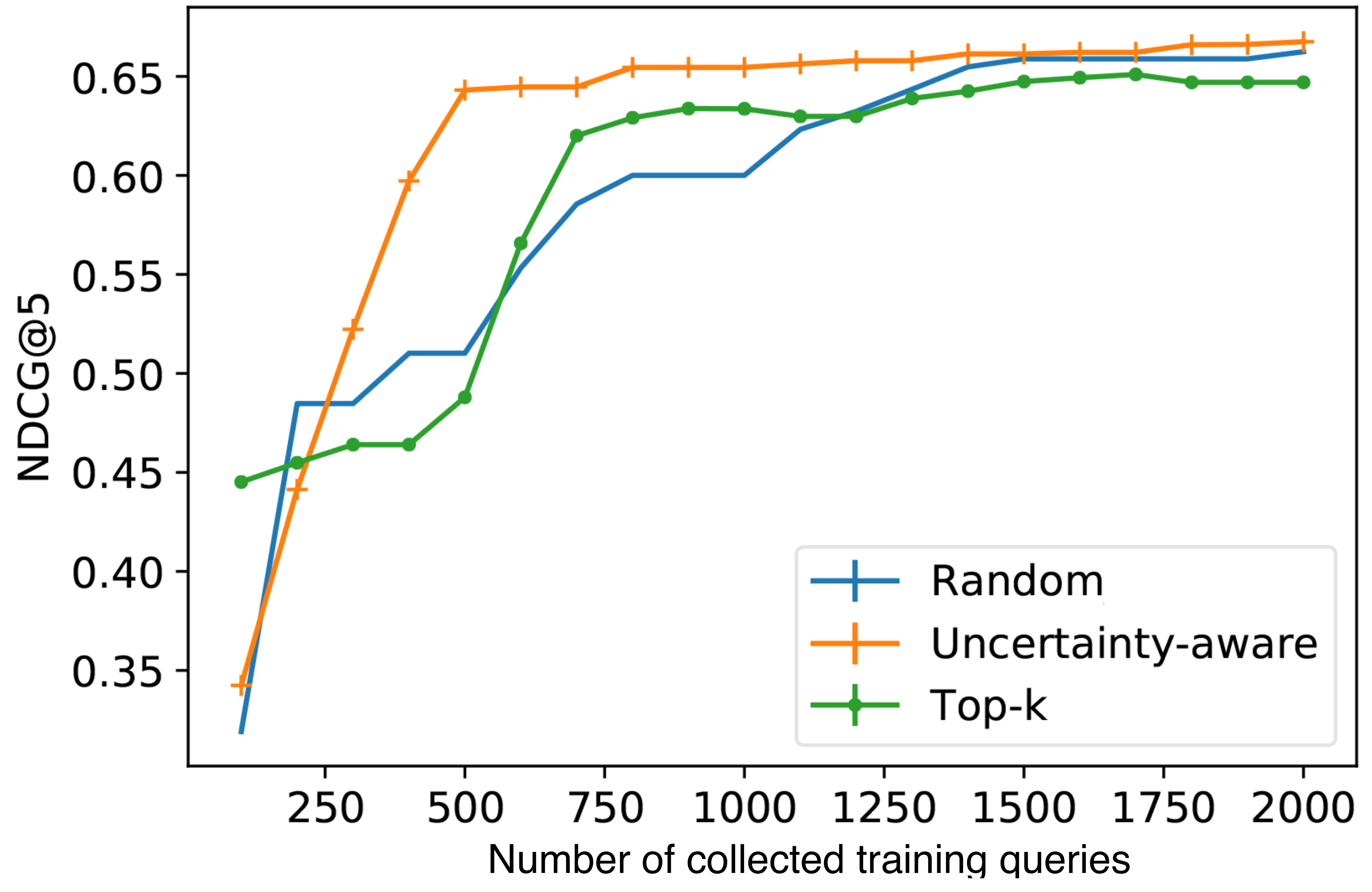}
	\caption{Performance of online learning to rank algorithms.}
	\label{fig:OLTR_uncertainty}
\end{figure}
In information theory~\cite{brillouin2013science}, information can be viewed as a resolution of uncertainty.
This indicates that an informative training example should be the data samples that can significantly reduce uncertainty in model outputs~\cite{settles2009active}.
Model uncertainty (or variance) has been extensively studied in explainable AI~\cite{bobek2021introducing} and accountable AI~\cite{carvalho2021assessing}.
Specifically in ranking optimization, however, it's potential hasn't been fully explored in existing literature.
Recently, studies~\cite{cohen2021not,yang2022can} on online learning to rank has shown that model uncertainty estimation could be helpful for identifying informative data for efficient training and model optimization.
To better illustrate this, we followed the experiment setup on MSLR-30K\footnote{https://www.microsoft.com/en-us/research/project/mslr/} used by Yang et al.~\cite{yang2022can} and compared the ranking functions learned with (1) Top-k: simply sort and show candidate results based on their relevance scores predicted by the current ranking model; (2) Random: randomly sort and present candidate results; (3) Uncertainty-aware: sort and show results based on their relevance scores plus the uncertainty (estimated with random dropout~\cite{gal2016dropout}) of current model outputs.
As shown in Figure~\ref{fig:OLTR_uncertainty}, by selecting and presenting items based on their ranking score uncertainty (i.e., the orange line), we can collect training data that significantly increase the convergence speed and ranking performance of a ranking algorithm.
This motivates us to further investigate the potential of uncertainty modeling for efficient metric-agnostic ranking optimization. 

Based on these observations, we propose to conduct uncertainty estimation for models in metric-agnostic ranking optimization and use the estimated uncertainty information to improve the efficiency of training data collection.
Our idea is to collect ground truth data on cases where the current ranking model or surrogate metric model has the highest uncertainty so that algorithms can converge quickly with less training data.
Formally, let the uncertainty/variance of a surrogate metric model $g(\pi)$ over ranking $\pi$ be $\delta_{\pi}$, and the uncertainty/variance of the probability to observe $\pi$ given the ranking model (i.e., $Pr(\pi|f,\theta)$ in Eq.~(\ref{equ:true_stochastic_loss}) or $\mathcal{H}(\{f(x|\theta)|x\in I\}|\rho)$ in Eq.~(\ref{equ:detached_ranking})) be $\epsilon_{pi}$.
For simplicity, we assume that $g(\pi)$ and $Pr(\pi)$ are independent variables following normal distribution $\mathcal{N}(g(\pi),\delta_{\pi}^2)$ and $\mathcal{N}(Pr(\pi),\epsilon_{\pi}^2)$.
Then the uncertainty of expected ranking reward in Eq.~(\ref{equ:true_stochastic_loss}) or (\ref{equ:scoring_ranking_optimization}), i.e., $\Delta(I)$, can be computed as 
\begin{equation}
\begin{split}
\Delta(I)^2 &= \sum_{\pi\in\mathcal{R}(I)}Var(g(\pi)\cdot Pr(\pi))^2 \\
&= \sum_{\pi\in\mathcal{R}(I)}(g(\pi)^2\epsilon_{\pi}^2 +Pr(\pi)^2\delta_{\pi}^2 +\epsilon_{\pi}^2\delta_{\pi}^2)
\end{split}
\label{equ:reward_uncertainty}
\end{equation} 
where $\mathcal{R}(I)$ is all possible rankings in session item set $I$.
Based on the theory of active learning~\cite{cohn1996active}, annotating data with highest model uncertainty could sharply decrease the number of training examples the learner needs in order to achieve a good performance.
Therefore, to conduct efficient optimization, we can actively select ranking session $I$ based on $\Delta(I)$ and select specific ranking $\pi\in\mathcal{R}(I)$ based on $Var(g(\pi)\cdot Pr(\pi))$ to collect ground truth metric score $m(\pi)$ in online experiments.
Finally, for the computation of $\delta_{\pi}$ and $\epsilon_{\pi}$, we could directly utilize variational inference~\cite{hoffman2013stochastic} and Monte Carlo dropout~\cite{gal2016dropout,kristiadi2020being,nair2010rectified} to infer them from the current surrogate metric models and ranking models.


\vspace{-8pt}
\subsection{Hypersurface and State Reorganization}
Existing ML theories and techniques are well-developed for convex optimization problems.
Particularly, gradient descent based optimizers~\cite{ruder2016overview,bubeck2017convex} have been shown to be effective in finding global optima for differentiable convex loss or reward functions.
When the loss or reward distribution is non-convex, however, there is no guarantee that those methods could produce good results and gradient descent could easily be stuck by local minimums.
Non-convex reward hypersurface is often an important problem for complex ranking metrics and it could jeopardize the efficiency and effectiveness of optimization algorithms.


A key reason behind the non-convex reward hypersurface in metric-agnostic ranking optimization is the fact that, for complex ranking metrics, the swap of two adjacent items does not guarantee a consistent gradient direction for the ranking reward.
Previously, under the PRP assumption, the change of relative positions for two items would always lead to a better or worse reward in terms of ranking metrics~\cite{robertson1976relevance}.
Thus, ranking losses built for these metrics are mostly convex and algorithms that directly optimize the partial orders of items could converge quickly while achieving reasonably good performance~\cite{liu2009learning}.
In contrast, complex ranking metrics have more complicated structures and could be affected by factors which may not be quantified or known in advance.
For example, user engagement in news recommendation is significantly affected by item novelty. 
The swap of two items could lead to both increase and decrease of user engagement depending on the context.
Without proper solutions, these phenomena naturally create non-convex reward hypersurface that are difficult to optimize.

\begin{figure}[t]
	\includegraphics[scale=0.37]{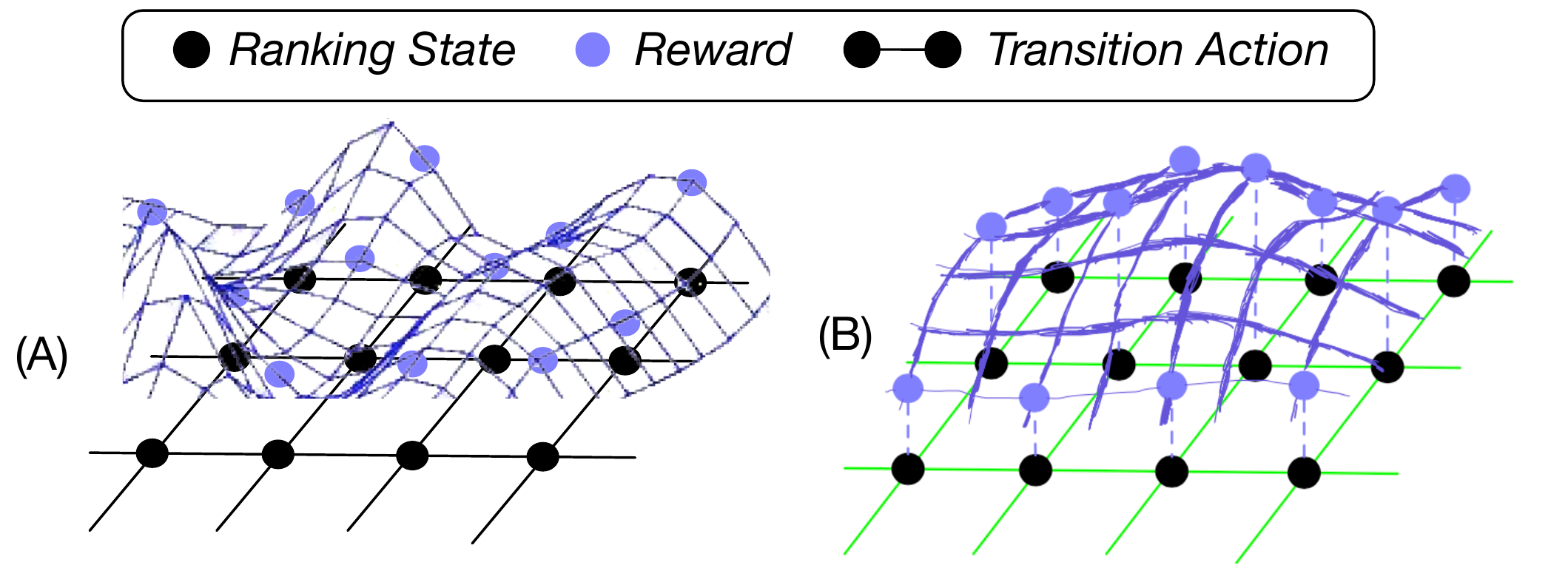} 
	\caption{An illustration of state reorganization.}
	\label{fig:state_reorg}
\end{figure}
Therefore, to solve the problem of non-convex reward hypersurface, we want to explore the possibility of refining the reward hypersurface of complex ranking metrics for better optimization efficiency and effectiveness.
We may achieve this through reward analysis and ranking state reorganization.
Similar to the ranking state formulation in Section~\ref{sec:random_flips}, suppose that we have already created a differentiable reward function with techniques proposed in Section~\ref{subsec-T2}, which give us a reward hypersurface over ranking states such as Figure~\ref{fig:state_reorg}(A).
When we connect ranking states naively through item flips (as discussed in Section~\ref{sec:random_flips}), the reward hypersurface of complex ranking metrics could be non-convex with respect to the changes of ranking states.
To solve this problem, one needs to conduct detailed analysis on the ranking rewards and group ranking states with similar properties together.
For example, we could conduct data clustering with both the reward scores and the local context features of each ranking, assuming that data within the same cluster would have more consistent characteristics with respect to parameter perturbations on ranking models.
Based on the result of reward analysis, we could develop new ranking paradigms that create a new set of transaction actions that connect states with similar properties intra and inter clusters to build a reward hypersurface with better convexity. 

The development of such ranking paradigms is directly related to the design of ranking module in Section~\ref{sec:detach}.
For example, instead of connecting ranking states with random item flips, we could build a ranking module that selects both positions and items with probabilistic models.
We could optimize the sampling distributions together with the surrogate reward model so that ranking states with similar predicted rewards would be neighbors under the new ranking paradigm. 
Intuitively, this means reorganizing ranking states through a new set of transaction actions so that we could create a reward hypersurface with less local optima, as shown in Figure~\ref{fig:state_reorg}(B).
Though this is a difficult task with non-trivial solutions, we believe that any progress on this direction could significantly contribute to the studies of complex ranking metric optimization and potentially change the future of ranking optimization research.

%% file: sec-related.tex
\vspace{-10pt}
\section{Related Work}
\label{sec-related}

\textbf{Learning to Rank and Ranking Optimization}.
Learning to rank (LTR) refers to the techniques of adapting ML algorithms to build ranking models for IR problems~\cite{liu2009learning}. 
The basic idea of LTR is to build feature representations for each item and design a scoring function that takes the features to predict the ranking score of each item~\cite{liu2009learning}.
Traditional LTR models usually assume the independence of item relevance~\cite{robertson1976relevance} and score each item purely based on its individual feature vector~\cite{liu2009learning}.
Recently, more advanced LTR techniques acknowledge the importance of contextual information and create multi-variate scoring functions that jointly score multiple items together with all their feature vectors~\cite{ai2018learning,jiang2018beyond,ai2019groupwise,pasumarthi2020permutation,pang2020setrank}.
To optimize the parameters in ranking models, classic LTR methods design different loss functions based on hand-crafted ranking metrics such as NDCG~\cite{wang2018lambdaloss}.
Depending how many items are concerned in each loss function, LTR loss functions can be broadly categorized into pointwise~\cite{gey1994inferring}, pairwise~\cite{joachims2006training,burges2005learning}, and listwise\cite{burges2006learning,cao2007learning} methods.
Despite their differences, existing loss functions require explicit knowledge on the ranking metrics and fine-grained labels on each item~\cite{liu2009learning}, which is often unavailable in metric-agnostic ranking optimization.

\textbf{Ranking Evaluation and Performance Metrics}.
Ranking evaluation is critical for IR as it essentially determines the direction of system optimization~\cite{cleverdon1967cranfield,voorhees2001philosophy}.
Classic ranking evaluation studies mostly focus on evaluating IR systems based on their ability to retrieve relevant documents.
For instance, a variety of ranking metrics are designed based on the relevance of each item~\cite{saracevic1996relevance} and use different heuristics to model the importance of different ranking positions~\cite{jarvelin2002cumulated,kelly2007effects,chapelle2009expected}.
In late 1990s, ranking evaluation becomes more complicated when the objective of IR moves from retrieving topical relevant documents to documents that satisfies the individual needs of search engine or recommendation users~\cite{verberne2013reliability}.
For example, document's relevance to users depending on multiple factors such as authority, timeliness, novelty, etc., many of which cannot be captured by the topical relevance between queries and documents~\cite{mao2016does}.
Thus, modern ranking evaluation on real IR systems often rely on user-centric and session-level metrics such as user engagements, user satisfactions, etc.~\cite{cole2009usefulness,liu2015different,odijk2015struggling}.
These metrics usually suffer from noisy user behaviors and cannot be explicitly formulated based on individual relevance labels. 
How to optimize these metrics directly is mostly unknown in existing literature.

\textbf{Unbiased and Online Learning to Rank}.
Studies on unbiased and online LTR can be viewed as some initial attempts to optimize ranking with noisy metrics. 
In general, user interactions with IR systems often implicitly reflect their opinions and satisfaction on result rankings~\cite{hassan2010beyond}.
Therefore, modern IR systems often use user behavior data (e.g., clicks) as relevance signals to improve ranking systems~\cite{joachims2002optimizing}.
As user behaviors often contain significant noise and collecting user interactions is expensive and time-consuming, studies on unbiased LTR and online LTR focus on how to effectively remove noise and bias from user data and efficiently train LTR models with less user interactions~\cite{ai2018ULR}.
Examples include the applications of counterfactual learning~\cite{wang2016learning,joachims2017unbiased,ai2018unbiased,agarwal2019addressing}, variance reductions techniques in bandit learning~\cite{schuth2016multileave,wang2018efficient,wang2019variance,jagerman2019model}, etc..
Nonetheless, existing studies on unbiased and online LTR mostly focus on how to interpret item-level user behavior signals and collect relevance feedback on each result~\cite{ai2021unbiased}.
From this perspective, they are similar to existing LTR methods using fine-grained relevance annotations. 

\textbf{Optimization Algorithms and ML Paradigms}.
In last decades, we have observed significant improvements in the design of optimization algorithms in ML~\cite{sra2012optimization,jain2017non,bottou2018optimization}.
Depending on the format of training data, existing ML methods can be broadly categorized as supervised, semi-supervised, unsupervised algorithms.
Also, depending on the format of supervision signals, a variety of learning algorithms have been proposed, included reinforcement learning~\cite{kaelbling1996reinforcement,sutton2018reinforcement}, few-shot learning~\cite{ravi2016optimization,garcia2017few,sung2018learning}, multi-task learning~\cite{evgeniou2004regularized,ruder2017overview}, adversarial learning~\cite{lowd2005adversarial,wang2017generative,zhang2018mitigating}, etc..
Despite the different formulations used by these algorithms, most of them optimize model parameters following the classic gradient descent framework by iteratively updating models with their current gradients with respect to the learning targets~\cite{bottou2010large}.

%% file: sec-broader.tex
\vspace{-10pt}
\section{Conclusion}
\label{sec-conclusion}

In this paper, we discuss the challenges and opportunities of metric-agnostic ranking optimization, and propose three research tasks, i.e., structure-agnostic metric modeling and simulation, differentiable ranking process with coarse-grained reward, and efficient ranking exploration in parameter space.
Given that ranking has served as a critical component of most IR systems and IR applications have become the most important tools for people to access information online, the development of effective techniques to support better ranking models and optimization algorithms will be significant. 
Besides, studies on metric-agnostic ranking optimization could have great value from a broader perspective of AI technology as advanced ranking optimization techniques could lead to a higher level of machine intelligence from passive learning to active learning in real-world environments. 
Through this paper, we hope to encourage more researchers to look into the problem and develop new ranking paradigms and optimization algorithms in future.






%% file: sigir2023.bbl

\begin{thebibliography}{103}


\ifx \showCODEN    \undefined \def \showCODEN     #1{\unskip}     \fi
\ifx \showDOI      \undefined \def \showDOI       #1{#1}\fi
\ifx \showISBNx    \undefined \def \showISBNx     #1{\unskip}     \fi
\ifx \showISBNxiii \undefined \def \showISBNxiii  #1{\unskip}     \fi
\ifx \showISSN     \undefined \def \showISSN      #1{\unskip}     \fi
\ifx \showLCCN     \undefined \def \showLCCN      #1{\unskip}     \fi
\ifx \shownote     \undefined \def \shownote      #1{#1}          \fi
\ifx \showarticletitle \undefined \def \showarticletitle #1{#1}   \fi
\ifx \showURL      \undefined \def \showURL       {\relax}        \fi
\providecommand\bibfield[2]{#2}
\providecommand\bibinfo[2]{#2}
\providecommand\natexlab[1]{#1}
\providecommand\showeprint[2][]{arXiv:#2}

\bibitem[Agarwal et~al\mbox{.}(2019)]%
        {agarwal2019addressing}
\bibfield{author}{\bibinfo{person}{Aman Agarwal}, \bibinfo{person}{Xuanhui
  Wang}, \bibinfo{person}{Cheng Li}, \bibinfo{person}{Michael Bendersky}, {and}
  \bibinfo{person}{Marc Najork}.} \bibinfo{year}{2019}\natexlab{}.
\newblock \showarticletitle{Addressing trust bias for unbiased
  learning-to-rank}. In \bibinfo{booktitle}{\emph{The World Wide Web
  Conference}}. \bibinfo{pages}{4--14}.
\newblock


\bibitem[Ai et~al\mbox{.}(2018a)]%
        {ai2018learning}
\bibfield{author}{\bibinfo{person}{Qingyao Ai}, \bibinfo{person}{Keping Bi},
  \bibinfo{person}{Jiafeng Guo}, {and} \bibinfo{person}{W~Bruce Croft}.}
  \bibinfo{year}{2018}\natexlab{a}.
\newblock \showarticletitle{Learning a deep listwise context model for ranking
  refinement}. In \bibinfo{booktitle}{\emph{The 41st International ACM SIGIR
  Conference on Research \& Development in Information Retrieval}}. ACM,
  \bibinfo{pages}{135--144}.
\newblock


\bibitem[Ai et~al\mbox{.}(2018b)]%
        {ai2018unbiased}
\bibfield{author}{\bibinfo{person}{Qingyao Ai}, \bibinfo{person}{Keping Bi},
  \bibinfo{person}{Cheng Luo}, \bibinfo{person}{Jiafeng Guo}, {and}
  \bibinfo{person}{W~Bruce Croft}.} \bibinfo{year}{2018}\natexlab{b}.
\newblock \showarticletitle{Unbiased learning to rank with unbiased propensity
  estimation}. In \bibinfo{booktitle}{\emph{The 41st International ACM SIGIR
  Conference on Research \& Development in Information Retrieval}}. ACM,
  \bibinfo{pages}{385--394}.
\newblock


\bibitem[Ai et~al\mbox{.}(2018c)]%
        {ai2018ULR}
\bibfield{author}{\bibinfo{person}{Qingyao Ai}, \bibinfo{person}{Jiaxin Mao},
  \bibinfo{person}{Yiqun Liu}, {and} \bibinfo{person}{W~Bruce Croft}.}
  \bibinfo{year}{2018}\natexlab{c}.
\newblock \showarticletitle{Unbiased learning to rank: Theory and practice}. In
  \bibinfo{booktitle}{\emph{Proceedings of the 27th ACM International
  Conference on Information and Knowledge Management}}. ACM,
  \bibinfo{pages}{2305--2306}.
\newblock


\bibitem[Ai et~al\mbox{.}(2019)]%
        {ai2019groupwise}
\bibfield{author}{\bibinfo{person}{Qingyao Ai}, \bibinfo{person}{Xuanhui Wang},
  \bibinfo{person}{Sebastian Bruch}, \bibinfo{person}{Nadav Golbandi},
  \bibinfo{person}{Michael Bendersky}, {and} \bibinfo{person}{Marc Najork}.}
  \bibinfo{year}{2019}\natexlab{}.
\newblock \showarticletitle{Learning Groupwise Multivariate Scoring Functions
  Using Deep Neural Networks}. In \bibinfo{booktitle}{\emph{Proceedings of the
  2019 ACM international conference on the theory of information retrieval}}.
  ACM.
\newblock


\bibitem[Ai et~al\mbox{.}(2021)]%
        {ai2021unbiased}
\bibfield{author}{\bibinfo{person}{Qingyao Ai}, \bibinfo{person}{Tao Yang},
  \bibinfo{person}{Huazheng Wang}, {and} \bibinfo{person}{Jiaxin Mao}.}
  \bibinfo{year}{2021}\natexlab{}.
\newblock \showarticletitle{Unbiased Learning to Rank: Online or Offline?}
\newblock \bibinfo{journal}{\emph{ACM Transactions on Information Systems
  (TOIS)}} \bibinfo{volume}{39}, \bibinfo{number}{2} (\bibinfo{year}{2021}),
  \bibinfo{pages}{1--29}.
\newblock


\bibitem[Al-Maskari et~al\mbox{.}(2007)]%
        {10.1145/1277741.1277902}
\bibfield{author}{\bibinfo{person}{Azzah Al-Maskari}, \bibinfo{person}{Mark
  Sanderson}, {and} \bibinfo{person}{Paul Clough}.}
  \bibinfo{year}{2007}\natexlab{}.
\newblock \showarticletitle{The Relationship between IR Effectiveness Measures
  and User Satisfaction}. In \bibinfo{booktitle}{\emph{Proceedings of the 30th
  Annual International ACM SIGIR Conference on Research and Development in
  Information Retrieval}} (Amsterdam, The Netherlands)
  \emph{(\bibinfo{series}{SIGIR '07})}. \bibinfo{publisher}{Association for
  Computing Machinery}, \bibinfo{address}{New York, NY, USA},
  \bibinfo{pages}{773–774}.
\newblock
\showISBNx{9781595935977}
\urldef\tempurl%
\url{https://doi.org/10.1145/1277741.1277902}
\showDOI{\tempurl}


\bibitem[Arabzadeh et~al\mbox{.}(2021)]%
        {arabzadeh2021shallow}
\bibfield{author}{\bibinfo{person}{Negar Arabzadeh}, \bibinfo{person}{Alexandra
  Vtyurina}, \bibinfo{person}{Xinyi Yan}, {and} \bibinfo{person}{Charles~LA
  Clarke}.} \bibinfo{year}{2021}\natexlab{}.
\newblock \showarticletitle{Shallow pooling for sparse labels}.
\newblock \bibinfo{journal}{\emph{arXiv preprint arXiv:2109.00062}}
  (\bibinfo{year}{2021}).
\newblock


\bibitem[Biega et~al\mbox{.}(2018)]%
        {biega2018equity}
\bibfield{author}{\bibinfo{person}{Asia~J Biega}, \bibinfo{person}{Krishna~P
  Gummadi}, {and} \bibinfo{person}{Gerhard Weikum}.}
  \bibinfo{year}{2018}\natexlab{}.
\newblock \showarticletitle{Equity of attention: Amortizing individual fairness
  in rankings}. In \bibinfo{booktitle}{\emph{The 41st international acm sigir
  conference on research \& development in information retrieval}}.
  \bibinfo{pages}{405--414}.
\newblock


\bibitem[Bobek and Nalepa(2021)]%
        {bobek2021introducing}
\bibfield{author}{\bibinfo{person}{Szymon Bobek} {and}
  \bibinfo{person}{Grzegorz~J Nalepa}.} \bibinfo{year}{2021}\natexlab{}.
\newblock \showarticletitle{Introducing Uncertainty into Explainable AI
  Methods}. In \bibinfo{booktitle}{\emph{International Conference on
  Computational Science}}. Springer, \bibinfo{pages}{444--457}.
\newblock


\bibitem[Bottou(2010)]%
        {bottou2010large}
\bibfield{author}{\bibinfo{person}{L{\'e}on Bottou}.}
  \bibinfo{year}{2010}\natexlab{}.
\newblock \showarticletitle{Large-scale machine learning with stochastic
  gradient descent}.
\newblock In \bibinfo{booktitle}{\emph{Proceedings of COMPSTAT'2010}}.
  \bibinfo{publisher}{Springer}, \bibinfo{pages}{177--186}.
\newblock


\bibitem[Bottou et~al\mbox{.}(2018)]%
        {bottou2018optimization}
\bibfield{author}{\bibinfo{person}{L{\'e}on Bottou}, \bibinfo{person}{Frank~E
  Curtis}, {and} \bibinfo{person}{Jorge Nocedal}.}
  \bibinfo{year}{2018}\natexlab{}.
\newblock \showarticletitle{Optimization methods for large-scale machine
  learning}.
\newblock \bibinfo{journal}{\emph{Siam Review}} \bibinfo{volume}{60},
  \bibinfo{number}{2} (\bibinfo{year}{2018}), \bibinfo{pages}{223--311}.
\newblock


\bibitem[Brillouin(2013)]%
        {brillouin2013science}
\bibfield{author}{\bibinfo{person}{Leon Brillouin}.}
  \bibinfo{year}{2013}\natexlab{}.
\newblock \bibinfo{booktitle}{\emph{Science and information theory}}.
\newblock \bibinfo{publisher}{Courier Corporation}.
\newblock


\bibitem[Bubeck(2017)]%
        {bubeck2017convex}
\bibfield{author}{\bibinfo{person}{S{\'e}bastien Bubeck}.}
  \bibinfo{year}{2017}\natexlab{}.
\newblock \showarticletitle{Convex Optimization: Algorithms and Complexity}.
\newblock \bibinfo{journal}{\emph{Foundations and Trends in Machine Learning}}
  \bibinfo{volume}{8} (\bibinfo{year}{2017}).
\newblock


\bibitem[Buckley et~al\mbox{.}(2007)]%
        {buckley2007bias}
\bibfield{author}{\bibinfo{person}{Chris Buckley}, \bibinfo{person}{Darrin
  Dimmick}, \bibinfo{person}{Ian Soboroff}, {and} \bibinfo{person}{Ellen
  Voorhees}.} \bibinfo{year}{2007}\natexlab{}.
\newblock \showarticletitle{Bias and the limits of pooling for large
  collections}.
\newblock \bibinfo{journal}{\emph{Information retrieval}} \bibinfo{volume}{10},
  \bibinfo{number}{6} (\bibinfo{year}{2007}), \bibinfo{pages}{491--508}.
\newblock


\bibitem[Buckman et~al\mbox{.}(2020)]%
        {buckman2020importance}
\bibfield{author}{\bibinfo{person}{Jacob Buckman}, \bibinfo{person}{Carles
  Gelada}, {and} \bibinfo{person}{Marc~G Bellemare}.}
  \bibinfo{year}{2020}\natexlab{}.
\newblock \showarticletitle{The importance of pessimism in fixed-dataset policy
  optimization}.
\newblock \bibinfo{journal}{\emph{arXiv preprint arXiv:2009.06799}}
  (\bibinfo{year}{2020}).
\newblock


\bibitem[Burges et~al\mbox{.}(2006)]%
        {burges2006learning}
\bibfield{author}{\bibinfo{person}{Christopher Burges}, \bibinfo{person}{Robert
  Ragno}, {and} \bibinfo{person}{Quoc Le}.} \bibinfo{year}{2006}\natexlab{}.
\newblock \showarticletitle{Learning to rank with nonsmooth cost functions}.
\newblock \bibinfo{journal}{\emph{Advances in neural information processing
  systems}}  \bibinfo{volume}{19} (\bibinfo{year}{2006}),
  \bibinfo{pages}{193--200}.
\newblock


\bibitem[Burges et~al\mbox{.}(2005)]%
        {burges2005learning}
\bibfield{author}{\bibinfo{person}{Chris Burges}, \bibinfo{person}{Tal Shaked},
  \bibinfo{person}{Erin Renshaw}, \bibinfo{person}{Ari Lazier},
  \bibinfo{person}{Matt Deeds}, \bibinfo{person}{Nicole Hamilton}, {and}
  \bibinfo{person}{Greg Hullender}.} \bibinfo{year}{2005}\natexlab{}.
\newblock \showarticletitle{Learning to rank using gradient descent}. In
  \bibinfo{booktitle}{\emph{Proceedings of the 22nd international conference on
  Machine learning}}. \bibinfo{pages}{89--96}.
\newblock


\bibitem[Cai et~al\mbox{.}(2022)]%
        {cai2022negative}
\bibfield{author}{\bibinfo{person}{Yinqiong Cai}, \bibinfo{person}{Jiafeng
  Guo}, \bibinfo{person}{Yixing Fan}, \bibinfo{person}{Qingyao Ai},
  \bibinfo{person}{Ruqing Zhang}, {and} \bibinfo{person}{Xueqi Cheng}.}
  \bibinfo{year}{2022}\natexlab{}.
\newblock \showarticletitle{Hard Negatives or False Negatives: Correcting
  Pooling Bias in Training Neural Ranking Models}. In
  \bibinfo{booktitle}{\emph{Proceedings of the 31st ACM International
  Conference on Information \& Knowledge Management}}.
  \bibinfo{pages}{118--127}.
\newblock


\bibitem[Cao et~al\mbox{.}(2007)]%
        {cao2007learning}
\bibfield{author}{\bibinfo{person}{Zhe Cao}, \bibinfo{person}{Tao Qin},
  \bibinfo{person}{Tie-Yan Liu}, \bibinfo{person}{Ming-Feng Tsai}, {and}
  \bibinfo{person}{Hang Li}.} \bibinfo{year}{2007}\natexlab{}.
\newblock \showarticletitle{Learning to rank: from pairwise approach to
  listwise approach}. In \bibinfo{booktitle}{\emph{Proceedings of the 24th
  international conference on Machine learning}}. \bibinfo{pages}{129--136}.
\newblock


\bibitem[Carvalho et~al\mbox{.}(2021)]%
        {carvalho2021assessing}
\bibfield{author}{\bibinfo{person}{Paula Carvalho}, \bibinfo{person}{Danielle
  Caled}, \bibinfo{person}{M{\'a}rio~J Silva}, \bibinfo{person}{Bruno Martins},
  \bibinfo{person}{Jo{\~a}o~Paulo Carvalho}, \bibinfo{person}{Joaquim
  Carreira}, \bibinfo{person}{Jo{\~a}o~Pedro Fonseca}, \bibinfo{person}{Teresa
  Gomes}, {and} \bibinfo{person}{Pedro Camacho}.}
  \bibinfo{year}{2021}\natexlab{}.
\newblock \showarticletitle{Assessing News Credibility: Misinformation Content
  Indicators}.
\newblock  (\bibinfo{year}{2021}).
\newblock


\bibitem[Chapelle et~al\mbox{.}(2009)]%
        {chapelle2009expected}
\bibfield{author}{\bibinfo{person}{Olivier Chapelle}, \bibinfo{person}{Donald
  Metlzer}, \bibinfo{person}{Ya Zhang}, {and} \bibinfo{person}{Pierre
  Grinspan}.} \bibinfo{year}{2009}\natexlab{}.
\newblock \showarticletitle{Expected reciprocal rank for graded relevance}. In
  \bibinfo{booktitle}{\emph{Proceedings of the 18th ACM conference on
  Information and knowledge management}}. \bibinfo{pages}{621--630}.
\newblock


\bibitem[Chen et~al\mbox{.}(2017)]%
        {chen2017user}
\bibfield{author}{\bibinfo{person}{Ye Chen}, \bibinfo{person}{Yiqun Liu},
  \bibinfo{person}{Min Zhang}, {and} \bibinfo{person}{Shaoping Ma}.}
  \bibinfo{year}{2017}\natexlab{}.
\newblock \showarticletitle{User satisfaction prediction with mouse movement
  information in heterogeneous search environment}.
\newblock \bibinfo{journal}{\emph{IEEE Transactions on Knowledge and Data
  Engineering}} \bibinfo{volume}{29}, \bibinfo{number}{11}
  (\bibinfo{year}{2017}), \bibinfo{pages}{2470--2483}.
\newblock


\bibitem[Chuklin et~al\mbox{.}(2013)]%
        {chuklin2013click}
\bibfield{author}{\bibinfo{person}{Aleksandr Chuklin}, \bibinfo{person}{Pavel
  Serdyukov}, {and} \bibinfo{person}{Maarten De~Rijke}.}
  \bibinfo{year}{2013}\natexlab{}.
\newblock \showarticletitle{Click model-based information retrieval metrics}.
  In \bibinfo{booktitle}{\emph{Proceedings of the 36th international ACM SIGIR
  conference on Research and development in information retrieval}}.
  \bibinfo{pages}{493--502}.
\newblock


\bibitem[Cleverdon(1967)]%
        {cleverdon1967cranfield}
\bibfield{author}{\bibinfo{person}{Cyril Cleverdon}.}
  \bibinfo{year}{1967}\natexlab{}.
\newblock \showarticletitle{The Cranfield tests on index language devices}. In
  \bibinfo{booktitle}{\emph{Aslib proceedings}}. MCB UP Ltd.
\newblock


\bibitem[Cohen et~al\mbox{.}(2021)]%
        {cohen2021not}
\bibfield{author}{\bibinfo{person}{Daniel Cohen}, \bibinfo{person}{Bhaskar
  Mitra}, \bibinfo{person}{Oleg Lesota}, \bibinfo{person}{Navid Rekabsaz},
  {and} \bibinfo{person}{Carsten Eickhoff}.} \bibinfo{year}{2021}\natexlab{}.
\newblock \showarticletitle{Not all relevance scores are equal: Efficient
  uncertainty and calibration modeling for deep retrieval models}. In
  \bibinfo{booktitle}{\emph{Proceedings of the 44th International ACM SIGIR
  Conference on Research and Development in Information Retrieval}}.
  \bibinfo{pages}{654--664}.
\newblock


\bibitem[Cohn et~al\mbox{.}(1996)]%
        {cohn1996active}
\bibfield{author}{\bibinfo{person}{David~A Cohn}, \bibinfo{person}{Zoubin
  Ghahramani}, {and} \bibinfo{person}{Michael~I Jordan}.}
  \bibinfo{year}{1996}\natexlab{}.
\newblock \showarticletitle{Active learning with statistical models}.
\newblock \bibinfo{journal}{\emph{Journal of artificial intelligence research}}
   \bibinfo{volume}{4} (\bibinfo{year}{1996}), \bibinfo{pages}{129--145}.
\newblock


\bibitem[Cole et~al\mbox{.}(2009)]%
        {cole2009usefulness}
\bibfield{author}{\bibinfo{person}{Michael Cole}, \bibinfo{person}{Jingjing
  Liu}, \bibinfo{person}{Nicholas Belkin}, \bibinfo{person}{Ralf Bierig},
  \bibinfo{person}{Jacek Gwizdka}, \bibinfo{person}{C Liu},
  \bibinfo{person}{Jin Zhang}, {and} \bibinfo{person}{X Zhang}.}
  \bibinfo{year}{2009}\natexlab{}.
\newblock \showarticletitle{Usefulness as the criterion for evaluation of
  interactive information retrieval}.
\newblock \bibinfo{journal}{\emph{Proc. HCIR}} (\bibinfo{year}{2009}),
  \bibinfo{pages}{1--4}.
\newblock


\bibitem[Craswell et~al\mbox{.}(2008)]%
        {craswell2008experimental}
\bibfield{author}{\bibinfo{person}{Nick Craswell}, \bibinfo{person}{Onno
  Zoeter}, \bibinfo{person}{Michael Taylor}, {and} \bibinfo{person}{Bill
  Ramsey}.} \bibinfo{year}{2008}\natexlab{}.
\newblock \showarticletitle{An experimental comparison of click position-bias
  models}. In \bibinfo{booktitle}{\emph{Proceedings of the 2008 international
  conference on web search and data mining}}. \bibinfo{pages}{87--94}.
\newblock


\bibitem[Dadashi et~al\mbox{.}(2021)]%
        {dadashi2021offline}
\bibfield{author}{\bibinfo{person}{Robert Dadashi}, \bibinfo{person}{Shideh
  Rezaeifar}, \bibinfo{person}{Nino Vieillard}, \bibinfo{person}{L{\'e}onard
  Hussenot}, \bibinfo{person}{Olivier Pietquin}, {and}
  \bibinfo{person}{Matthieu Geist}.} \bibinfo{year}{2021}\natexlab{}.
\newblock \showarticletitle{Offline Reinforcement Learning with Pseudometric
  Learning}.
\newblock \bibinfo{journal}{\emph{arXiv preprint arXiv:2103.01948}}
  (\bibinfo{year}{2021}).
\newblock


\bibitem[Dehghani et~al\mbox{.}(2017)]%
        {dehghani2017neural}
\bibfield{author}{\bibinfo{person}{Mostafa Dehghani}, \bibinfo{person}{Hamed
  Zamani}, \bibinfo{person}{Aliaksei Severyn}, \bibinfo{person}{Jaap Kamps},
  {and} \bibinfo{person}{W~Bruce Croft}.} \bibinfo{year}{2017}\natexlab{}.
\newblock \showarticletitle{Neural ranking models with weak supervision}. In
  \bibinfo{booktitle}{\emph{Proceedings of the 40th International ACM SIGIR
  Conference on Research and Development in Information Retrieval}}.
  \bibinfo{pages}{65--74}.
\newblock


\bibitem[Devlin et~al\mbox{.}(2019)]%
        {devlin2019bert}
\bibfield{author}{\bibinfo{person}{Jacob Devlin}, \bibinfo{person}{Ming-Wei
  Chang}, \bibinfo{person}{Kenton Lee}, {and} \bibinfo{person}{Kristina
  Toutanova}.} \bibinfo{year}{2019}\natexlab{}.
\newblock \showarticletitle{BERT: Pre-training of Deep Bidirectional
  Transformers for Language Understanding}. In
  \bibinfo{booktitle}{\emph{NAACL-HLT (1)}}.
\newblock


\bibitem[Eiron and McCurley(2003)]%
        {eiron2003analysis}
\bibfield{author}{\bibinfo{person}{Nadav Eiron} {and} \bibinfo{person}{Kevin~S
  McCurley}.} \bibinfo{year}{2003}\natexlab{}.
\newblock \showarticletitle{Analysis of anchor text for web search}. In
  \bibinfo{booktitle}{\emph{Proceedings of the 26th annual international ACM
  SIGIR conference on Research and development in informaion retrieval}}.
  \bibinfo{pages}{459--460}.
\newblock


\bibitem[Evgeniou and Pontil(2004)]%
        {evgeniou2004regularized}
\bibfield{author}{\bibinfo{person}{Theodoros Evgeniou} {and}
  \bibinfo{person}{Massimiliano Pontil}.} \bibinfo{year}{2004}\natexlab{}.
\newblock \showarticletitle{Regularized multi--task learning}. In
  \bibinfo{booktitle}{\emph{Proceedings of the tenth ACM SIGKDD international
  conference on Knowledge discovery and data mining}}.
  \bibinfo{pages}{109--117}.
\newblock


\bibitem[Gal and Ghahramani(2016)]%
        {gal2016dropout}
\bibfield{author}{\bibinfo{person}{Yarin Gal} {and} \bibinfo{person}{Zoubin
  Ghahramani}.} \bibinfo{year}{2016}\natexlab{}.
\newblock \showarticletitle{Dropout as a bayesian approximation: Representing
  model uncertainty in deep learning}. In
  \bibinfo{booktitle}{\emph{international conference on machine learning}}.
  PMLR, \bibinfo{pages}{1050--1059}.
\newblock


\bibitem[Garcia and Bruna(2017)]%
        {garcia2017few}
\bibfield{author}{\bibinfo{person}{Victor Garcia} {and} \bibinfo{person}{Joan
  Bruna}.} \bibinfo{year}{2017}\natexlab{}.
\newblock \showarticletitle{Few-shot learning with graph neural networks}.
\newblock \bibinfo{journal}{\emph{arXiv preprint arXiv:1711.04043}}
  (\bibinfo{year}{2017}).
\newblock


\bibitem[Gey(1994)]%
        {gey1994inferring}
\bibfield{author}{\bibinfo{person}{Fredric~C Gey}.}
  \bibinfo{year}{1994}\natexlab{}.
\newblock \showarticletitle{Inferring probability of relevance using the method
  of logistic regression}. In \bibinfo{booktitle}{\emph{SIGIR’94}}. Springer,
  \bibinfo{pages}{222--231}.
\newblock


\bibitem[Goodfellow et~al\mbox{.}(2016)]%
        {goodfellow2016deep}
\bibfield{author}{\bibinfo{person}{Ian Goodfellow}, \bibinfo{person}{Yoshua
  Bengio}, {and} \bibinfo{person}{Aaron Courville}.}
  \bibinfo{year}{2016}\natexlab{}.
\newblock \bibinfo{booktitle}{\emph{Deep learning}}.
\newblock \bibinfo{publisher}{MIT press}.
\newblock


\bibitem[Hassan et~al\mbox{.}(2010)]%
        {hassan2010beyond}
\bibfield{author}{\bibinfo{person}{Ahmed Hassan}, \bibinfo{person}{Rosie
  Jones}, {and} \bibinfo{person}{Kristina~Lisa Klinkner}.}
  \bibinfo{year}{2010}\natexlab{}.
\newblock \showarticletitle{Beyond DCG: user behavior as a predictor of a
  successful search}. In \bibinfo{booktitle}{\emph{Proceedings of the third ACM
  international conference on Web search and data mining}}.
  \bibinfo{pages}{221--230}.
\newblock


\bibitem[Hauff et~al\mbox{.}(2009)]%
        {hauff2009combination}
\bibfield{author}{\bibinfo{person}{Claudia Hauff}, \bibinfo{person}{Leif
  Azzopardi}, {and} \bibinfo{person}{Djoerd Hiemstra}.}
  \bibinfo{year}{2009}\natexlab{}.
\newblock \showarticletitle{The combination and evaluation of query performance
  prediction methods}. In \bibinfo{booktitle}{\emph{European Conference on
  Information Retrieval}}. Springer, \bibinfo{pages}{301--312}.
\newblock


\bibitem[Hoffman et~al\mbox{.}(2013)]%
        {hoffman2013stochastic}
\bibfield{author}{\bibinfo{person}{Matthew~D Hoffman}, \bibinfo{person}{David~M
  Blei}, \bibinfo{person}{Chong Wang}, {and} \bibinfo{person}{John Paisley}.}
  \bibinfo{year}{2013}\natexlab{}.
\newblock \showarticletitle{Stochastic variational inference.}
\newblock \bibinfo{journal}{\emph{Journal of Machine Learning Research}}
  \bibinfo{volume}{14}, \bibinfo{number}{5} (\bibinfo{year}{2013}).
\newblock


\bibitem[Jagerman et~al\mbox{.}(2019)]%
        {jagerman2019model}
\bibfield{author}{\bibinfo{person}{Rolf Jagerman}, \bibinfo{person}{Harrie
  Oosterhuis}, {and} \bibinfo{person}{Maarten de Rijke}.}
  \bibinfo{year}{2019}\natexlab{}.
\newblock \showarticletitle{To model or to intervene: A comparison of
  counterfactual and online learning to rank from user interactions}. In
  \bibinfo{booktitle}{\emph{Proceedings of the 42nd international ACM SIGIR
  conference on research and development in information retrieval}}.
  \bibinfo{pages}{15--24}.
\newblock


\bibitem[Jain and Kar(2017)]%
        {jain2017non}
\bibfield{author}{\bibinfo{person}{Prateek Jain} {and}
  \bibinfo{person}{Purushottam Kar}.} \bibinfo{year}{2017}\natexlab{}.
\newblock \showarticletitle{Non-convex optimization for machine learning}.
\newblock \bibinfo{journal}{\emph{arXiv preprint arXiv:1712.07897}}
  (\bibinfo{year}{2017}).
\newblock


\bibitem[J{\"a}rvelin and Kek{\"a}l{\"a}inen(2002)]%
        {jarvelin2002cumulated}
\bibfield{author}{\bibinfo{person}{Kalervo J{\"a}rvelin} {and}
  \bibinfo{person}{Jaana Kek{\"a}l{\"a}inen}.} \bibinfo{year}{2002}\natexlab{}.
\newblock \showarticletitle{Cumulated gain-based evaluation of IR techniques}.
\newblock \bibinfo{journal}{\emph{ACM Transactions on Information Systems
  (TOIS)}} \bibinfo{volume}{20}, \bibinfo{number}{4} (\bibinfo{year}{2002}),
  \bibinfo{pages}{422--446}.
\newblock


\bibitem[Jiang and Allan(2016)]%
        {10.1145/2854946.2855005}
\bibfield{author}{\bibinfo{person}{Jiepu Jiang} {and} \bibinfo{person}{James
  Allan}.} \bibinfo{year}{2016}\natexlab{}.
\newblock \showarticletitle{Correlation Between System and User Metrics in a
  Session}. In \bibinfo{booktitle}{\emph{Proceedings of the 2016 ACM on
  Conference on Human Information Interaction and Retrieval}} (Carrboro, North
  Carolina, USA) \emph{(\bibinfo{series}{CHIIR '16})}.
  \bibinfo{publisher}{Association for Computing Machinery},
  \bibinfo{address}{New York, NY, USA}, \bibinfo{pages}{285–288}.
\newblock
\showISBNx{9781450337519}
\urldef\tempurl%
\url{https://doi.org/10.1145/2854946.2855005}
\showDOI{\tempurl}


\bibitem[Jiang et~al\mbox{.}(2018)]%
        {jiang2018beyond}
\bibfield{author}{\bibinfo{person}{Ray Jiang}, \bibinfo{person}{Sven Gowal},
  \bibinfo{person}{Timothy~A Mann}, {and} \bibinfo{person}{Danilo~J Rezende}.}
  \bibinfo{year}{2018}\natexlab{}.
\newblock \showarticletitle{Beyond greedy ranking: Slate optimization via
  list-CVAE}.
\newblock \bibinfo{journal}{\emph{arXiv preprint arXiv:1803.01682}}
  (\bibinfo{year}{2018}).
\newblock


\bibitem[Joachims(2002)]%
        {joachims2002optimizing}
\bibfield{author}{\bibinfo{person}{Thorsten Joachims}.}
  \bibinfo{year}{2002}\natexlab{}.
\newblock \showarticletitle{Optimizing search engines using clickthrough data}.
  In \bibinfo{booktitle}{\emph{Proceedings of the eighth ACM SIGKDD
  international conference on Knowledge discovery and data mining}}.
  \bibinfo{pages}{133--142}.
\newblock


\bibitem[Joachims(2006)]%
        {joachims2006training}
\bibfield{author}{\bibinfo{person}{Thorsten Joachims}.}
  \bibinfo{year}{2006}\natexlab{}.
\newblock \showarticletitle{Training linear SVMs in linear time}. In
  \bibinfo{booktitle}{\emph{Proceedings of the 12th ACM SIGKDD international
  conference on Knowledge discovery and data mining}}.
  \bibinfo{pages}{217--226}.
\newblock


\bibitem[Joachims et~al\mbox{.}(2017)]%
        {joachims2017unbiased}
\bibfield{author}{\bibinfo{person}{Thorsten Joachims}, \bibinfo{person}{Adith
  Swaminathan}, {and} \bibinfo{person}{Tobias Schnabel}.}
  \bibinfo{year}{2017}\natexlab{}.
\newblock \showarticletitle{Unbiased learning-to-rank with biased feedback}. In
  \bibinfo{booktitle}{\emph{Proceedings of the Tenth ACM International
  Conference on Web Search and Data Mining}}. \bibinfo{pages}{781--789}.
\newblock


\bibitem[Kaelbling et~al\mbox{.}(1996)]%
        {kaelbling1996reinforcement}
\bibfield{author}{\bibinfo{person}{Leslie~Pack Kaelbling},
  \bibinfo{person}{Michael~L Littman}, {and} \bibinfo{person}{Andrew~W Moore}.}
  \bibinfo{year}{1996}\natexlab{}.
\newblock \showarticletitle{Reinforcement learning: A survey}.
\newblock \bibinfo{journal}{\emph{Journal of artificial intelligence research}}
   \bibinfo{volume}{4} (\bibinfo{year}{1996}), \bibinfo{pages}{237--285}.
\newblock


\bibitem[Kelly et~al\mbox{.}(2007)]%
        {kelly2007effects}
\bibfield{author}{\bibinfo{person}{Diane Kelly}, \bibinfo{person}{Xin Fu},
  {and} \bibinfo{person}{Chirag Shah}.} \bibinfo{year}{2007}\natexlab{}.
\newblock \showarticletitle{Effects of rank and precision of search results on
  users’ evaluations of system performance}.
\newblock \bibinfo{journal}{\emph{University of North Carolina}}
  (\bibinfo{year}{2007}).
\newblock


\bibitem[Khandelwal et~al\mbox{.}(2021)]%
        {khandelwal2021jointly}
\bibfield{author}{\bibinfo{person}{Hitesh Khandelwal}, \bibinfo{person}{Viet
  Ha-Thuc}, \bibinfo{person}{Avishek Dutta}, \bibinfo{person}{Yining Lu},
  \bibinfo{person}{Nan Du}, \bibinfo{person}{Zhihao Li}, {and}
  \bibinfo{person}{Qi Hu}.} \bibinfo{year}{2021}\natexlab{}.
\newblock \showarticletitle{Jointly Optimize Capacity, Latency and Engagement
  in Large-scale Recommendation Systems}. In
  \bibinfo{booktitle}{\emph{Fifteenth ACM Conference on Recommender Systems}}.
  \bibinfo{pages}{559--561}.
\newblock


\bibitem[Kristiadi et~al\mbox{.}(2020)]%
        {kristiadi2020being}
\bibfield{author}{\bibinfo{person}{Agustinus Kristiadi},
  \bibinfo{person}{Matthias Hein}, {and} \bibinfo{person}{Philipp Hennig}.}
  \bibinfo{year}{2020}\natexlab{}.
\newblock \showarticletitle{Being bayesian, even just a bit, fixes
  overconfidence in relu networks}. In \bibinfo{booktitle}{\emph{International
  Conference on Machine Learning}}. PMLR, \bibinfo{pages}{5436--5446}.
\newblock


\bibitem[Lagun et~al\mbox{.}(2014)]%
        {lagun2014towards}
\bibfield{author}{\bibinfo{person}{Dmitry Lagun}, \bibinfo{person}{Chih-Hung
  Hsieh}, \bibinfo{person}{Dale Webster}, {and} \bibinfo{person}{Vidhya
  Navalpakkam}.} \bibinfo{year}{2014}\natexlab{}.
\newblock \showarticletitle{Towards better measurement of attention and
  satisfaction in mobile search}. In \bibinfo{booktitle}{\emph{Proceedings of
  the 37th international ACM SIGIR conference on Research \& development in
  information retrieval}}. \bibinfo{pages}{113--122}.
\newblock


\bibitem[Liu et~al\mbox{.}(2018)]%
        {liu2018satisfaction}
\bibfield{author}{\bibinfo{person}{Mengyang Liu}, \bibinfo{person}{Yiqun Liu},
  \bibinfo{person}{Jiaxin Mao}, \bibinfo{person}{Cheng Luo},
  \bibinfo{person}{Min Zhang}, {and} \bibinfo{person}{Shaoping Ma}.}
  \bibinfo{year}{2018}\natexlab{}.
\newblock \showarticletitle{" Satisfaction with Failure" or" Unsatisfied
  Success" Investigating the Relationship between Search Success and User
  Satisfaction}. In \bibinfo{booktitle}{\emph{Proceedings of the 2018 World
  Wide Web Conference}}. \bibinfo{pages}{1533--1542}.
\newblock


\bibitem[Liu(2009)]%
        {liu2009learning}
\bibfield{author}{\bibinfo{person}{Tie-Yan Liu}.}
  \bibinfo{year}{2009}\natexlab{}.
\newblock \showarticletitle{Learning to rank for information retrieval}.
\newblock \bibinfo{journal}{\emph{Foundations and Trends in Information
  Retrieval}} \bibinfo{volume}{3}, \bibinfo{number}{3} (\bibinfo{year}{2009}),
  \bibinfo{pages}{225--331}.
\newblock


\bibitem[Liu et~al\mbox{.}(2015)]%
        {liu2015different}
\bibfield{author}{\bibinfo{person}{Yiqun Liu}, \bibinfo{person}{Ye Chen},
  \bibinfo{person}{Jinhui Tang}, \bibinfo{person}{Jiashen Sun},
  \bibinfo{person}{Min Zhang}, \bibinfo{person}{Shaoping Ma}, {and}
  \bibinfo{person}{Xuan Zhu}.} \bibinfo{year}{2015}\natexlab{}.
\newblock \showarticletitle{Different users, different opinions: Predicting
  search satisfaction with mouse movement information}. In
  \bibinfo{booktitle}{\emph{Proceedings of the 38th international ACM SIGIR
  conference on research and development in information retrieval}}.
  \bibinfo{pages}{493--502}.
\newblock


\bibitem[Lowd and Meek(2005)]%
        {lowd2005adversarial}
\bibfield{author}{\bibinfo{person}{Daniel Lowd} {and}
  \bibinfo{person}{Christopher Meek}.} \bibinfo{year}{2005}\natexlab{}.
\newblock \showarticletitle{Adversarial learning}. In
  \bibinfo{booktitle}{\emph{Proceedings of the eleventh ACM SIGKDD
  international conference on Knowledge discovery in data mining}}.
  \bibinfo{pages}{641--647}.
\newblock


\bibitem[Ma et~al\mbox{.}(2021)]%
        {ma2021learning}
\bibfield{author}{\bibinfo{person}{Jiaqi Ma}, \bibinfo{person}{Xinyang Yi},
  \bibinfo{person}{Weijing Tang}, \bibinfo{person}{Zhe Zhao},
  \bibinfo{person}{Lichan Hong}, \bibinfo{person}{Ed Chi}, {and}
  \bibinfo{person}{Qiaozhu Mei}.} \bibinfo{year}{2021}\natexlab{}.
\newblock \showarticletitle{Learning-to-Rank with Partitioned Preference: Fast
  Estimation for the Plackett-Luce Model}. In
  \bibinfo{booktitle}{\emph{International Conference on Artificial Intelligence
  and Statistics}}. PMLR, \bibinfo{pages}{928--936}.
\newblock


\bibitem[Mao et~al\mbox{.}(2016)]%
        {mao2016does}
\bibfield{author}{\bibinfo{person}{Jiaxin Mao}, \bibinfo{person}{Yiqun Liu},
  \bibinfo{person}{Ke Zhou}, \bibinfo{person}{Jian-Yun Nie},
  \bibinfo{person}{Jingtao Song}, \bibinfo{person}{Min Zhang},
  \bibinfo{person}{Shaoping Ma}, \bibinfo{person}{Jiashen Sun}, {and}
  \bibinfo{person}{Hengliang Luo}.} \bibinfo{year}{2016}\natexlab{}.
\newblock \showarticletitle{When does relevance mean usefulness and user
  satisfaction in web search?}. In \bibinfo{booktitle}{\emph{Proceedings of the
  39th International ACM SIGIR conference on Research and Development in
  Information Retrieval}}. \bibinfo{pages}{463--472}.
\newblock


\bibitem[Metzler and Croft(2005)]%
        {metzler2005markov}
\bibfield{author}{\bibinfo{person}{Donald Metzler} {and}
  \bibinfo{person}{W~Bruce Croft}.} \bibinfo{year}{2005}\natexlab{}.
\newblock \showarticletitle{A markov random field model for term dependencies}.
  In \bibinfo{booktitle}{\emph{Proceedings of the 28th annual international ACM
  SIGIR conference on Research and development in information retrieval}}.
  \bibinfo{pages}{472--479}.
\newblock


\bibitem[Morik et~al\mbox{.}(2020)]%
        {morik2020controlling}
\bibfield{author}{\bibinfo{person}{Marco Morik}, \bibinfo{person}{Ashudeep
  Singh}, \bibinfo{person}{Jessica Hong}, {and} \bibinfo{person}{Thorsten
  Joachims}.} \bibinfo{year}{2020}\natexlab{}.
\newblock \showarticletitle{Controlling fairness and bias in dynamic
  learning-to-rank}. In \bibinfo{booktitle}{\emph{Proceedings of the 43rd
  International ACM SIGIR Conference on Research and Development in Information
  Retrieval}}. \bibinfo{pages}{429--438}.
\newblock


\bibitem[Nair and Hinton(2010)]%
        {nair2010rectified}
\bibfield{author}{\bibinfo{person}{Vinod Nair} {and}
  \bibinfo{person}{Geoffrey~E Hinton}.} \bibinfo{year}{2010}\natexlab{}.
\newblock \showarticletitle{Rectified linear units improve restricted boltzmann
  machines}. In \bibinfo{booktitle}{\emph{Icml}}.
\newblock


\bibitem[Odijk et~al\mbox{.}(2015)]%
        {odijk2015struggling}
\bibfield{author}{\bibinfo{person}{Daan Odijk}, \bibinfo{person}{Ryen~W White},
  \bibinfo{person}{Ahmed Hassan~Awadallah}, {and} \bibinfo{person}{Susan~T
  Dumais}.} \bibinfo{year}{2015}\natexlab{}.
\newblock \showarticletitle{Struggling and success in web search}. In
  \bibinfo{booktitle}{\emph{Proceedings of the 24th ACM International on
  Conference on Information and Knowledge Management}}.
  \bibinfo{pages}{1551--1560}.
\newblock


\bibitem[Oosterhuis(2021)]%
        {oosterhuis2021computationally}
\bibfield{author}{\bibinfo{person}{Harrie Oosterhuis}.}
  \bibinfo{year}{2021}\natexlab{}.
\newblock \showarticletitle{Computationally Efficient Optimization of
  Plackett-Luce Ranking Models for Relevance and Fairness}.
\newblock \bibinfo{journal}{\emph{arXiv preprint arXiv:2105.00855}}
  (\bibinfo{year}{2021}).
\newblock


\bibitem[Oosterhuis and de~Rijke(2018)]%
        {oosterhuis2018differentiable}
\bibfield{author}{\bibinfo{person}{Harrie Oosterhuis} {and}
  \bibinfo{person}{Maarten de Rijke}.} \bibinfo{year}{2018}\natexlab{}.
\newblock \showarticletitle{Differentiable unbiased online learning to rank}.
  In \bibinfo{booktitle}{\emph{Proceedings of the 27th ACM International
  Conference on Information and Knowledge Management}}.
  \bibinfo{pages}{1293--1302}.
\newblock


\bibitem[Orabona and P{\'a}l(2016)]%
        {orabona2016parameter}
\bibfield{author}{\bibinfo{person}{Francesco Orabona} {and}
  \bibinfo{person}{D{\'a}vid P{\'a}l}.} \bibinfo{year}{2016}\natexlab{}.
\newblock \showarticletitle{Parameter-Free Convex Learning through Coin
  Betting}. In \bibinfo{booktitle}{\emph{Workshop on Automatic Machine
  Learning}}. PMLR, \bibinfo{pages}{75--82}.
\newblock


\bibitem[Pang et~al\mbox{.}(2020)]%
        {pang2020setrank}
\bibfield{author}{\bibinfo{person}{Liang Pang}, \bibinfo{person}{Jun Xu},
  \bibinfo{person}{Qingyao Ai}, \bibinfo{person}{Yanyan Lan},
  \bibinfo{person}{Xueqi Cheng}, {and} \bibinfo{person}{Jirong Wen}.}
  \bibinfo{year}{2020}\natexlab{}.
\newblock \showarticletitle{Setrank: Learning a permutation-invariant ranking
  model for information retrieval}. In \bibinfo{booktitle}{\emph{Proceedings of
  the 43rd International ACM SIGIR Conference on Research and Development in
  Information Retrieval}}. \bibinfo{pages}{499--508}.
\newblock


\bibitem[Pasumarthi et~al\mbox{.}(2020)]%
        {pasumarthi2020permutation}
\bibfield{author}{\bibinfo{person}{Rama~Kumar Pasumarthi},
  \bibinfo{person}{Honglei Zhuang}, \bibinfo{person}{Xuanhui Wang},
  \bibinfo{person}{Michael Bendersky}, {and} \bibinfo{person}{Marc Najork}.}
  \bibinfo{year}{2020}\natexlab{}.
\newblock \showarticletitle{Permutation equivariant document interaction
  network for neural learning to rank}. In
  \bibinfo{booktitle}{\emph{Proceedings of the 2020 ACM SIGIR on International
  Conference on Theory of Information Retrieval}}. \bibinfo{pages}{145--148}.
\newblock


\bibitem[Ravi and Larochelle(2016)]%
        {ravi2016optimization}
\bibfield{author}{\bibinfo{person}{Sachin Ravi} {and} \bibinfo{person}{Hugo
  Larochelle}.} \bibinfo{year}{2016}\natexlab{}.
\newblock \showarticletitle{Optimization as a model for few-shot learning}.
\newblock  (\bibinfo{year}{2016}).
\newblock


\bibitem[Robertson and Jones(1976)]%
        {robertson1976relevance}
\bibfield{author}{\bibinfo{person}{Stephen~E Robertson} {and}
  \bibinfo{person}{K~Sparck Jones}.} \bibinfo{year}{1976}\natexlab{}.
\newblock \showarticletitle{Relevance weighting of search terms}.
\newblock \bibinfo{journal}{\emph{Journal of the American Society for
  Information science}} \bibinfo{volume}{27}, \bibinfo{number}{3}
  (\bibinfo{year}{1976}), \bibinfo{pages}{129--146}.
\newblock


\bibitem[Ruder(2016)]%
        {ruder2016overview}
\bibfield{author}{\bibinfo{person}{Sebastian Ruder}.}
  \bibinfo{year}{2016}\natexlab{}.
\newblock \showarticletitle{An overview of gradient descent optimization
  algorithms}.
\newblock \bibinfo{journal}{\emph{arXiv preprint arXiv:1609.04747}}
  (\bibinfo{year}{2016}).
\newblock


\bibitem[Ruder(2017)]%
        {ruder2017overview}
\bibfield{author}{\bibinfo{person}{Sebastian Ruder}.}
  \bibinfo{year}{2017}\natexlab{}.
\newblock \showarticletitle{An overview of multi-task learning in deep neural
  networks}.
\newblock \bibinfo{journal}{\emph{arXiv preprint arXiv:1706.05098}}
  (\bibinfo{year}{2017}).
\newblock


\bibitem[Rue and Held(2005)]%
        {rue2005gaussian}
\bibfield{author}{\bibinfo{person}{Havard Rue} {and} \bibinfo{person}{Leonhard
  Held}.} \bibinfo{year}{2005}\natexlab{}.
\newblock \bibinfo{booktitle}{\emph{Gaussian Markov random fields: theory and
  applications}}.
\newblock \bibinfo{publisher}{CRC press}.
\newblock


\bibitem[Saracevic(1996)]%
        {saracevic1996relevance}
\bibfield{author}{\bibinfo{person}{Tefko Saracevic}.}
  \bibinfo{year}{1996}\natexlab{}.
\newblock \showarticletitle{Relevance reconsidered}. In
  \bibinfo{booktitle}{\emph{Proceedings of the second conference on conceptions
  of library and information science (CoLIS 2)}}. ACM New York,
  \bibinfo{pages}{201--218}.
\newblock


\bibitem[Schuth et~al\mbox{.}(2016)]%
        {schuth2016multileave}
\bibfield{author}{\bibinfo{person}{Anne Schuth}, \bibinfo{person}{Harrie
  Oosterhuis}, \bibinfo{person}{Shimon Whiteson}, {and}
  \bibinfo{person}{Maarten de Rijke}.} \bibinfo{year}{2016}\natexlab{}.
\newblock \showarticletitle{Multileave gradient descent for fast online
  learning to rank}. In \bibinfo{booktitle}{\emph{Proceedings of the Ninth ACM
  International Conference on Web Search and Data Mining}}.
  \bibinfo{pages}{457--466}.
\newblock


\bibitem[Seshadri et~al\mbox{.}(2020)]%
        {seshadri2020learning}
\bibfield{author}{\bibinfo{person}{Arjun Seshadri}, \bibinfo{person}{Stephen
  Ragain}, {and} \bibinfo{person}{Johan Ugander}.}
  \bibinfo{year}{2020}\natexlab{}.
\newblock \showarticletitle{Learning Rich Rankings}.
\newblock \bibinfo{journal}{\emph{Advances in Neural Information Processing
  Systems}} (\bibinfo{year}{2020}).
\newblock


\bibitem[Settles(2009)]%
        {settles2009active}
\bibfield{author}{\bibinfo{person}{Burr Settles}.}
  \bibinfo{year}{2009}\natexlab{}.
\newblock \showarticletitle{Active learning literature survey}.
\newblock  (\bibinfo{year}{2009}).
\newblock


\bibitem[Singh and Joachims(2018)]%
        {singh2018fairness}
\bibfield{author}{\bibinfo{person}{Ashudeep Singh} {and}
  \bibinfo{person}{Thorsten Joachims}.} \bibinfo{year}{2018}\natexlab{}.
\newblock \showarticletitle{Fairness of exposure in rankings}. In
  \bibinfo{booktitle}{\emph{Proceedings of the 24th ACM SIGKDD International
  Conference on Knowledge Discovery \& Data Mining}}.
  \bibinfo{pages}{2219--2228}.
\newblock


\bibitem[Sra et~al\mbox{.}(2012)]%
        {sra2012optimization}
\bibfield{author}{\bibinfo{person}{Suvrit Sra}, \bibinfo{person}{Sebastian
  Nowozin}, {and} \bibinfo{person}{Stephen~J Wright}.}
  \bibinfo{year}{2012}\natexlab{}.
\newblock \bibinfo{booktitle}{\emph{Optimization for machine learning}}.
\newblock \bibinfo{publisher}{Mit Press}.
\newblock


\bibitem[Sung et~al\mbox{.}(2018)]%
        {sung2018learning}
\bibfield{author}{\bibinfo{person}{Flood Sung}, \bibinfo{person}{Yongxin Yang},
  \bibinfo{person}{Li Zhang}, \bibinfo{person}{Tao Xiang},
  \bibinfo{person}{Philip~HS Torr}, {and} \bibinfo{person}{Timothy~M
  Hospedales}.} \bibinfo{year}{2018}\natexlab{}.
\newblock \showarticletitle{Learning to compare: Relation network for few-shot
  learning}. In \bibinfo{booktitle}{\emph{Proceedings of the IEEE conference on
  computer vision and pattern recognition}}. \bibinfo{pages}{1199--1208}.
\newblock


\bibitem[Sutton and Barto(2018)]%
        {sutton2018reinforcement}
\bibfield{author}{\bibinfo{person}{Richard~S Sutton} {and}
  \bibinfo{person}{Andrew~G Barto}.} \bibinfo{year}{2018}\natexlab{}.
\newblock \bibinfo{booktitle}{\emph{Reinforcement learning: An introduction}}.
\newblock \bibinfo{publisher}{MIT press}.
\newblock


\bibitem[Tkachenko and Lauw(2016)]%
        {tkachenko2016plackett}
\bibfield{author}{\bibinfo{person}{Maksim Tkachenko} {and}
  \bibinfo{person}{Hady~W Lauw}.} \bibinfo{year}{2016}\natexlab{}.
\newblock \showarticletitle{Plackett-luce regression mixture model for
  heterogeneous rankings}. In \bibinfo{booktitle}{\emph{Proceedings of the 25th
  ACM International on Conference on Information and Knowledge Management}}.
  \bibinfo{pages}{237--246}.
\newblock


\bibitem[Van~Laarhoven and Aarts(1987)]%
        {van1987simulated}
\bibfield{author}{\bibinfo{person}{Peter~JM Van~Laarhoven} {and}
  \bibinfo{person}{Emile~HL Aarts}.} \bibinfo{year}{1987}\natexlab{}.
\newblock \showarticletitle{Simulated annealing}.
\newblock In \bibinfo{booktitle}{\emph{Simulated annealing: Theory and
  applications}}. \bibinfo{publisher}{Springer}, \bibinfo{pages}{7--15}.
\newblock


\bibitem[Vaswani et~al\mbox{.}(2017)]%
        {vaswani2017attention}
\bibfield{author}{\bibinfo{person}{Ashish Vaswani}, \bibinfo{person}{Noam
  Shazeer}, \bibinfo{person}{Niki Parmar}, \bibinfo{person}{Jakob Uszkoreit},
  \bibinfo{person}{Llion Jones}, \bibinfo{person}{Aidan~N Gomez},
  \bibinfo{person}{{\L}ukasz Kaiser}, {and} \bibinfo{person}{Illia
  Polosukhin}.} \bibinfo{year}{2017}\natexlab{}.
\newblock \showarticletitle{Attention is all you need}. In
  \bibinfo{booktitle}{\emph{Advances in neural information processing
  systems}}. \bibinfo{pages}{5998--6008}.
\newblock


\bibitem[Verberne et~al\mbox{.}(2013)]%
        {verberne2013reliability}
\bibfield{author}{\bibinfo{person}{Suzan Verberne}, \bibinfo{person}{Maarten
  van~der Heijden}, \bibinfo{person}{Max Hinne}, \bibinfo{person}{Maya
  Sappelli}, \bibinfo{person}{Saskia Koldijk}, \bibinfo{person}{Eduard
  Hoenkamp}, {and} \bibinfo{person}{Wessel Kraaij}.}
  \bibinfo{year}{2013}\natexlab{}.
\newblock \showarticletitle{Reliability and validity of query intent
  assessments}.
\newblock \bibinfo{journal}{\emph{Journal of the American Society for
  Information Science and Technology}} \bibinfo{volume}{64},
  \bibinfo{number}{11} (\bibinfo{year}{2013}), \bibinfo{pages}{2224--2237}.
\newblock


\bibitem[Voorhees(2001)]%
        {voorhees2001philosophy}
\bibfield{author}{\bibinfo{person}{Ellen~M Voorhees}.}
  \bibinfo{year}{2001}\natexlab{}.
\newblock \showarticletitle{The philosophy of information retrieval
  evaluation}. In \bibinfo{booktitle}{\emph{Workshop of the cross-language
  evaluation forum for european languages}}. Springer,
  \bibinfo{pages}{355--370}.
\newblock


\bibitem[Wang et~al\mbox{.}(2019)]%
        {wang2019variance}
\bibfield{author}{\bibinfo{person}{Huazheng Wang}, \bibinfo{person}{Sonwoo
  Kim}, \bibinfo{person}{Eric McCord-Snook}, \bibinfo{person}{Qingyun Wu},
  {and} \bibinfo{person}{Hongning Wang}.} \bibinfo{year}{2019}\natexlab{}.
\newblock \showarticletitle{Variance reduction in gradient exploration for
  online learning to rank}. In \bibinfo{booktitle}{\emph{Proceedings of the
  42nd International ACM SIGIR Conference on Research and Development in
  Information Retrieval}}. \bibinfo{pages}{835--844}.
\newblock


\bibitem[Wang et~al\mbox{.}(2018a)]%
        {wang2018efficient}
\bibfield{author}{\bibinfo{person}{Huazheng Wang}, \bibinfo{person}{Ramsey
  Langley}, \bibinfo{person}{Sonwoo Kim}, \bibinfo{person}{Eric McCord-Snook},
  {and} \bibinfo{person}{Hongning Wang}.} \bibinfo{year}{2018}\natexlab{a}.
\newblock \showarticletitle{Efficient exploration of gradient space for online
  learning to rank}. In \bibinfo{booktitle}{\emph{The 41st International ACM
  SIGIR Conference on Research \& Development in Information Retrieval}}.
  \bibinfo{pages}{145--154}.
\newblock


\bibitem[Wang et~al\mbox{.}(2017)]%
        {wang2017generative}
\bibfield{author}{\bibinfo{person}{Kunfeng Wang}, \bibinfo{person}{Chao Gou},
  \bibinfo{person}{Yanjie Duan}, \bibinfo{person}{Yilun Lin},
  \bibinfo{person}{Xinhu Zheng}, {and} \bibinfo{person}{Fei-Yue Wang}.}
  \bibinfo{year}{2017}\natexlab{}.
\newblock \showarticletitle{Generative adversarial networks: introduction and
  outlook}.
\newblock \bibinfo{journal}{\emph{IEEE/CAA Journal of Automatica Sinica}}
  \bibinfo{volume}{4}, \bibinfo{number}{4} (\bibinfo{year}{2017}),
  \bibinfo{pages}{588--598}.
\newblock


\bibitem[Wang et~al\mbox{.}(2016)]%
        {wang2016learning}
\bibfield{author}{\bibinfo{person}{Xuanhui Wang}, \bibinfo{person}{Michael
  Bendersky}, \bibinfo{person}{Donald Metzler}, {and} \bibinfo{person}{Marc
  Najork}.} \bibinfo{year}{2016}\natexlab{}.
\newblock \showarticletitle{Learning to rank with selection bias in personal
  search}. In \bibinfo{booktitle}{\emph{Proceedings of the 39th International
  ACM SIGIR conference on Research and Development in Information Retrieval}}.
  \bibinfo{pages}{115--124}.
\newblock


\bibitem[Wang et~al\mbox{.}(2018b)]%
        {wang2018lambdaloss}
\bibfield{author}{\bibinfo{person}{Xuanhui Wang}, \bibinfo{person}{Cheng Li},
  \bibinfo{person}{Nadav Golbandi}, \bibinfo{person}{Michael Bendersky}, {and}
  \bibinfo{person}{Marc Najork}.} \bibinfo{year}{2018}\natexlab{b}.
\newblock \showarticletitle{The lambdaloss framework for ranking metric
  optimization}. In \bibinfo{booktitle}{\emph{Proceedings of the 27th ACM
  International Conference on Information and Knowledge Management}}.
  \bibinfo{pages}{1313--1322}.
\newblock


\bibitem[Wu et~al\mbox{.}(2021)]%
        {wu2021uncertainty}
\bibfield{author}{\bibinfo{person}{Yue Wu}, \bibinfo{person}{Shuangfei Zhai},
  \bibinfo{person}{Nitish Srivastava}, \bibinfo{person}{Joshua Susskind},
  \bibinfo{person}{Jian Zhang}, \bibinfo{person}{Ruslan Salakhutdinov}, {and}
  \bibinfo{person}{Hanlin Goh}.} \bibinfo{year}{2021}\natexlab{}.
\newblock \showarticletitle{Uncertainty Weighted Actor-Critic for Offline
  Reinforcement Learning}.
\newblock \bibinfo{journal}{\emph{arXiv preprint arXiv:2105.08140}}
  (\bibinfo{year}{2021}).
\newblock


\bibitem[Xia et~al\mbox{.}(2008)]%
        {xia2008listwise}
\bibfield{author}{\bibinfo{person}{Fen Xia}, \bibinfo{person}{Tie-Yan Liu},
  \bibinfo{person}{Jue Wang}, \bibinfo{person}{Wensheng Zhang}, {and}
  \bibinfo{person}{Hang Li}.} \bibinfo{year}{2008}\natexlab{}.
\newblock \showarticletitle{Listwise approach to learning to rank: theory and
  algorithm}. In \bibinfo{booktitle}{\emph{Proceedings of the 25th
  international conference on Machine learning}}. \bibinfo{pages}{1192--1199}.
\newblock


\bibitem[Xiao and Wang(2021)]%
        {xiao2021general}
\bibfield{author}{\bibinfo{person}{Teng Xiao} {and} \bibinfo{person}{Donglin
  Wang}.} \bibinfo{year}{2021}\natexlab{}.
\newblock \showarticletitle{A general offline reinforcement learning framework
  for interactive recommendation}. In \bibinfo{booktitle}{\emph{The
  Thirty-Fifth AAAI Conference on Artificial Intelligence, AAAI 2021}}.
\newblock


\bibitem[Yang et~al\mbox{.}(2020)]%
        {yang2020analysis}
\bibfield{author}{\bibinfo{person}{Tao Yang}, \bibinfo{person}{Shikai Fang},
  \bibinfo{person}{Shibo Li}, \bibinfo{person}{Yulan Wang}, {and}
  \bibinfo{person}{Qingyao Ai}.} \bibinfo{year}{2020}\natexlab{}.
\newblock \showarticletitle{Analysis of multivariate scoring functions for
  automatic unbiased learning to rank}. In
  \bibinfo{booktitle}{\emph{Proceedings of the 29th ACM International
  Conference on Information \& Knowledge Management}}.
  \bibinfo{pages}{2277--2280}.
\newblock


\bibitem[Yang et~al\mbox{.}(2022)]%
        {yang2022can}
\bibfield{author}{\bibinfo{person}{Tao Yang}, \bibinfo{person}{Chen Luo},
  \bibinfo{person}{Hanqing Lu}, \bibinfo{person}{Parth Gupta},
  \bibinfo{person}{Bing Yin}, {and} \bibinfo{person}{Qingyao Ai}.}
  \bibinfo{year}{2022}\natexlab{}.
\newblock \showarticletitle{Can clicks be both labels and features? Unbiased
  Behavior Feature Collection and Uncertainty-aware Learning to Rank}. In
  \bibinfo{booktitle}{\emph{Proceedings of the 45th International ACM SIGIR
  Conference on Research and Development in Information Retrieval}}.
  \bibinfo{pages}{6--17}.
\newblock


\bibitem[Yilmaz et~al\mbox{.}(2014)]%
        {yilmaz2014relevance}
\bibfield{author}{\bibinfo{person}{Emine Yilmaz}, \bibinfo{person}{Manisha
  Verma}, \bibinfo{person}{Nick Craswell}, \bibinfo{person}{Filip Radlinski},
  {and} \bibinfo{person}{Peter Bailey}.} \bibinfo{year}{2014}\natexlab{}.
\newblock \showarticletitle{Relevance and effort: An analysis of document
  utility}. In \bibinfo{booktitle}{\emph{Proceedings of the 23rd ACM
  International Conference on Conference on Information and Knowledge
  Management}}. \bibinfo{pages}{91--100}.
\newblock


\bibitem[Yue et~al\mbox{.}(2010)]%
        {yue2010beyond}
\bibfield{author}{\bibinfo{person}{Yisong Yue}, \bibinfo{person}{Rajan Patel},
  {and} \bibinfo{person}{Hein Roehrig}.} \bibinfo{year}{2010}\natexlab{}.
\newblock \showarticletitle{Beyond position bias: Examining result
  attractiveness as a source of presentation bias in clickthrough data}. In
  \bibinfo{booktitle}{\emph{Proceedings of the 19th international conference on
  World wide web}}. \bibinfo{pages}{1011--1018}.
\newblock


\bibitem[Zhang et~al\mbox{.}(2018)]%
        {zhang2018mitigating}
\bibfield{author}{\bibinfo{person}{Brian~Hu Zhang}, \bibinfo{person}{Blake
  Lemoine}, {and} \bibinfo{person}{Margaret Mitchell}.}
  \bibinfo{year}{2018}\natexlab{}.
\newblock \showarticletitle{Mitigating unwanted biases with adversarial
  learning}. In \bibinfo{booktitle}{\emph{Proceedings of the 2018 AAAI/ACM
  Conference on AI, Ethics, and Society}}. \bibinfo{pages}{335--340}.
\newblock


\bibitem[Zhao et~al\mbox{.}(2008)]%
        {zhao2008effective}
\bibfield{author}{\bibinfo{person}{Ying Zhao}, \bibinfo{person}{Falk Scholer},
  {and} \bibinfo{person}{Yohannes Tsegay}.} \bibinfo{year}{2008}\natexlab{}.
\newblock \showarticletitle{Effective pre-retrieval query performance
  prediction using similarity and variability evidence}. In
  \bibinfo{booktitle}{\emph{European conference on information retrieval}}.
  Springer, \bibinfo{pages}{52--64}.
\newblock


\bibitem[Zhao et~al\mbox{.}(2020)]%
        {zhao2020maximizing}
\bibfield{author}{\bibinfo{person}{Yifei Zhao}, \bibinfo{person}{Yu-Hang Zhou},
  \bibinfo{person}{Mingdong Ou}, \bibinfo{person}{Huan Xu}, {and}
  \bibinfo{person}{Nan Li}.} \bibinfo{year}{2020}\natexlab{}.
\newblock \showarticletitle{Maximizing Cumulative User Engagement in Sequential
  Recommendation: An Online Optimization Perspective}. In
  \bibinfo{booktitle}{\emph{Proceedings of the 26th ACM SIGKDD International
  Conference on Knowledge Discovery \& Data Mining}}.
  \bibinfo{pages}{2784--2792}.
\newblock


\bibitem[Zhou and Croft(2007)]%
        {zhou2007query}
\bibfield{author}{\bibinfo{person}{Yun Zhou} {and} \bibinfo{person}{W~Bruce
  Croft}.} \bibinfo{year}{2007}\natexlab{}.
\newblock \showarticletitle{Query performance prediction in web search
  environments}. In \bibinfo{booktitle}{\emph{Proceedings of the 30th annual
  international ACM SIGIR conference on Research and development in information
  retrieval}}. \bibinfo{pages}{543--550}.
\newblock


\end{thebibliography}
